\documentclass[11pt]{aastex}
\usepackage{emulateapj5,apjfonts}
\usepackage{onecolfloat}
\usepackage{epsf}
\ifx\epsfannounce\undefined \def\epsfannounce{\immediate\write16}\fi
 \epsfannounce{This is `epsf.tex' v2.7k <10 July 1997>}%
\newread\epsffilein    
\newif\ifepsfatend     
\newif\ifepsfbbfound   
\newif\ifepsfdraft     
\newif\ifepsffileok    
\newif\ifepsfframe     
\newif\ifepsfshow      
\epsfshowtrue          
\newif\ifepsfshowfilename 
\newif\ifepsfverbose   
\newdimen\epsfframemargin 
\newdimen\epsfframethickness 
\newdimen\epsfrsize    
\newdimen\epsftmp      
\newdimen\epsftsize    
\newdimen\epsfxsize    
\newdimen\epsfysize    
\newdimen\pspoints     
\def\epsfbox#1{\global\def\epsfllx{72}\global\def\epsflly{72}%
   \global\def\epsfurx{540}\global\def\epsfury{720}%
   \def\lbracket{[}\def\testit{#1}\ifx\testit\lbracket
   \let\next=\epsfgetlitbb\else\let\next=\epsfnormal\fi\next{#1}}%
\def\epsfgetlitbb#1#2 #3 #4 #5]#6{%
   \epsfgrab #2 #3 #4 #5 .\\%
   \epsfsetsize
   \epsfstatus{#6}%
   \epsfsetgraph{#6}%
}%
\def\epsfnormal#1{%
    \epsfgetbb{#1}%
    \epsfsetgraph{#1}%
}%
\newhelp\epsfnoopenhelp{The PostScript image file must be findable by
TeX, i.e., somewhere in the TEXINPUTS (or equivalent) path.}%
\def\epsfgetbb#1{%
    \openin\epsffilein=#1
    \ifeof\epsffilein
        \errhelp = \epsfnoopenhelp
        \errmessage{Could not open file #1, ignoring it}%
    \else                       
        {
            \chardef\other=12
            \def\do##1{\catcode`##1=\other}%
            \dospecials
            \catcode`\ =10
            \epsffileoktrue         
            \epsfatendfalse     
            \loop               
                \read\epsffilein to \epsffileline
                \ifeof\epsffilein 
                \epsffileokfalse 
            \else                
                \expandafter\epsfaux\epsffileline:. \\%
            \fi
            \ifepsffileok
            \repeat
            \ifepsfbbfound
            \else
                \ifepsfverbose
                    \immediate\write16{No BoundingBox comment found in %
                                    file #1; using defaults}%
                \fi
            \fi
        }
        \closein\epsffilein
    \fi                         
    \epsfsetsize                
    \epsfstatus{#1}%
}%
\def\epsfclipoff{\def\epsfclipstring{\ifepsfdraft\space clip\fi}}%
\epsfclipoff 
\def\epsfspecial#1{%
     \epsftmp=10\epsfxsize
     \divide\epsftmp\pspoints
     \ifnum\epsfrsize=0\relax
       \includegraphics{\ifepsfdraft}%
     \else
       \epsfrsize=10\epsfysize
       \divide\epsfrsize\pspoints
       \includegraphics{\ifepsfdraft}%
     \fi
}%
\def\epsfframe#1%
{%
  \leavevmode                   
  \setbox0 = \hbox{#1}%
  \dimen0 = \wd0                                
  \advance \dimen0 by 2\epsfframemargin         
  \advance \dimen0 by 2\epsfframethickness      
  \vbox
  {%
    \hrule height \epsfframethickness depth 0pt
    \hbox to \dimen0
    {%
      \hss
      \vrule width \epsfframethickness
      \kern \epsfframemargin
      \vbox {\kern \epsfframemargin \box0 \kern \epsfframemargin }%
      \kern \epsfframemargin
      \vrule width \epsfframethickness
      \hss
    }
    \hrule height 0pt depth \epsfframethickness
  }
}%
\def\epsfsetgraph#1%
{%
   \leavevmode
   \hbox{
     \ifepsfframe\expandafter\epsfframe\fi
     {\vbox to\epsfysize
     {%
        \ifepsfshow
            \vfil
            \hbox to \epsfxsize{\epsfspecial{#1}\hfil}%
        \else
            \vfil
            \hbox to\epsfxsize{%
               \hss
               \ifepsfshowfilename
               {%
                  \epsfframemargin=3pt 
                  \epsfframe{{\tt #1}}%
               }%
               \fi
               \hss
            }%
            \vfil
        \fi
     }%
   }}%
   \global\epsfxsize=0pt
   \global\epsfysize=0pt
}%
\def\epsfsetsize
{%
   \epsfrsize=\epsfury\pspoints
   \advance\epsfrsize by-\epsflly\pspoints
   \epsftsize=\epsfurx\pspoints
   \advance\epsftsize by-\epsfllx\pspoints
   \epsfxsize=\epsfsize{\epsftsize}{\epsfrsize}%
   \ifnum \epsfxsize=0
      \ifnum \epsfysize=0
        \epsfxsize=\epsftsize
        \epsfysize=\epsfrsize
        \epsfrsize=0pt
      \else
        \epsftmp=\epsftsize \divide\epsftmp\epsfrsize
        \epsfxsize=\epsfysize \multiply\epsfxsize\epsftmp
        \multiply\epsftmp\epsfrsize \advance\epsftsize-\epsftmp
        \epsftmp=\epsfysize
        \loop \advance\epsftsize\epsftsize \divide\epsftmp 2
        \ifnum \epsftmp>0
           \ifnum \epsftsize<\epsfrsize
           \else
              \advance\epsftsize-\epsfrsize \advance\epsfxsize\epsftmp
           \fi
        \repeat
        \epsfrsize=0pt
      \fi
   \else
     \ifnum \epsfysize=0
       \epsftmp=\epsfrsize \divide\epsftmp\epsftsize
       \epsfysize=\epsfxsize \multiply\epsfysize\epsftmp
       \multiply\epsftmp\epsftsize \advance\epsfrsize-\epsftmp
       \epsftmp=\epsfxsize
       \loop \advance\epsfrsize\epsfrsize \divide\epsftmp 2
       \ifnum \epsftmp>0
          \ifnum \epsfrsize<\epsftsize
          \else
             \advance\epsfrsize-\epsftsize \advance\epsfysize\epsftmp
          \fi
       \repeat
       \epsfrsize=0pt
     \else
       \epsfrsize=\epsfysize
     \fi
   \fi
}%
\def\epsfstatus#1{
   \ifepsfverbose
     \immediate\write16{#1: BoundingBox:
                  llx = \epsfllx\space lly = \epsflly\space
                  urx = \epsfurx\space ury = \epsfury\space}%
     \immediate\write16{#1: scaled width = \the\epsfxsize\space
                  scaled height = \the\epsfysize}%
   \fi
}%
{\catcode`\%=12 \global\let\epsfpercent=
\global\def\epsfatend{(atend)}%
\long\def\epsfaux#1#2:#3\\%
{%
   \def\testit{#2}
   \ifx#1\epsfpercent           
       \ifx\testit\epsfbblit    
            \epsfgrab #3 . . . \\%
            \ifx\epsfllx\epsfatend 
                \global\epsfatendtrue
            \else               
                \ifepsfatend    
                \else           
                    \epsffileokfalse
                \fi
                \global\epsfbbfoundtrue
            \fi
       \fi
   \fi
}%
\def\epsfempty{}%
\def\epsfgrab #1 #2 #3 #4 #5\\{%
   \global\def\epsfllx{#1}\ifx\epsfllx\epsfempty
      \epsfgrab #2 #3 #4 #5 .\\\else
   \global\def\epsflly{#2}%
   \global\def\epsfurx{#3}\global\def\epsfury{#4}\fi
}%
\def\epsfsize#1#2{\epsfxsize}%
\shorttitle{Tidally-Triggered Star Formation}
\shortauthors{Barton Gillespie et al.}

\newcommand{\ew}  {{EW(H$\alpha$)}}
\newcommand{\aew}  {{$\left<{\rm EW(H\alpha)}\right>$}}

\newcommand{\ewb} {{EW(H$\alpha$)$_{\rm burst}$({\rm t})}}
\newcommand{\ewm} {{EW(H$\alpha$)$_{\rm meas}$({\rm t})}}
\newcommand{\fbo} {{f$_{\rm B0}$}}

\newcommand{\fbn} {{f$_{\rm Bn}$({\rm t})}}
\newcommand{\fro} {{f$_{\rm R0}$}}

\newcommand{\frn} {{f$_{\rm Rn}$({\rm t})}}
\newcommand{\BRs} {{$(B-R)_{\rm s}$}}
\newcommand{\aBRs}{{$\left<(B-R)_{\rm s}\right>$}}
\newcommand{\BRn} {{$(B-R)_{\rm n}$}}
\newcommand{\BRo} {{$(B-R)_{0}$}}
\newcommand{\kf} {{{\rm k}$(z,{\rm SED_{0}, SED_{b}})$}}
\newcommand{\co}  {{C$_0$}}

\newcommand{\cn}  {{C$_{\rm n}$({\rm t})}}

\newcommand{\strr}  {{s$_{\rm R}$({\rm t})}}
\newcommand{\stro}  {{s$_{\rm 100}$}}
\newcommand{\dd}    {{$\Delta D$}}
\newcommand{\Vc}   {{V$^{\rm c}_{2.2}$}}
\newcommand{\mew}  {{{\rm EW}({\rm H}\alpha)}}
\newcommand{\maew} {{\left<{\rm EW(H\alpha)}\right>}}
\newcommand{\maBRs}{{\left<(B-R)_{\rm s}\right>}}
\newcommand{\mewm} {{{\rm EW}({\rm H}\alpha)_{\rm meas}({\rm t})}}
\newcommand{\mewb} {{{\rm EW}({\rm H}\alpha)_{\rm burst}({\rm t})}}
\newcommand{\mfbo} {{f_{\rm B0}}}

\newcommand{\mfbn} {{f_{\rm Bn}({\rm t})}}
\newcommand{\mfro} {{f_{\rm R0}}}

\newcommand{\mfrn} {{f_{\rm Rn}({\rm t})}}
\newcommand{\mkf}   {{{\rm k}(z,{\rm SED_{0}, SED_{b}})}}
\newcommand{\mco}  {{{\rm C}_0}}
\newcommand{\mcs}  {{{\rm C}_{\rm s}({\rm t})}}
\newcommand{\mcn}  {{{\rm C}_{\rm n}({\rm t})}}

\newcommand{\msr}  {{{\rm s}_{\rm R}({\rm t})}}
\newcommand{\mstr}  {{{\rm s}_{\rm R}({\rm t})}}
\newcommand{\mso}  {{{\rm s}_{\rm 100}}}
\newcommand{\mBR}  {{(B-R)_{\rm s}}}
\newcommand{\mBRs}  {{(B-R)_{\rm s}}}
\newcommand{\mBRn} {{(B-R)_{\rm n}}}
\newcommand{\mBRo} {{(B-R)_{0}}}

\newcommand{\mVc}  {{{{\rm V}^{\rm c}_{2.2}}}}

\begin{document}

\def\head{

\title{Tidally-Triggered Star Formation in Close Pairs of Galaxies 2: 
Constraints on Burst Strengths and Ages}
\author{Elizabeth Barton Gillespie\altaffilmark{1}}
\affil{University of Arizona, Steward Observatory, 933 N. Cherry Ave.,
Tucson, AZ 85721 (email: bgillespie@as.arizona.edu)}
\and \author{Margaret J. Geller, Scott J. Kenyon}
\affil{Smithsonian Astrophysical Observatory,
60 Garden St., Cambridge, MA 02138
(email: mgeller,skenyon@cfa.harvard.edu)}

\begin{abstract}

Galaxy-galaxy interactions rearrange the baryons in galaxies
and trigger substantial star formation;
the aggregate effects of these interactions on the evolutionary
histories of galaxies in the Universe are poorly understood.
We combine $B$ and $R$-band photometry and optical spectroscopy
to estimate the strengths and timescales of bursts of
triggered star formation in the centers of 190 galaxies in pairs
and compact groups.  Based on an analysis of the measured
colors and \ew, we characterize the pre-existing and triggered
populations separately.  The best-fitting burst scenarios assume 
stronger reddening corrections for line emission than for
the continuum and continuous star formation lasting for 
$\gtrsim$ a hundred Myr.  The most realistic scenarios require an 
initial mass function that is deficient in the highest-mass stars.
The color of the pre-existing stellar population is the most significant 
source of uncertainty.

Triggered star formation contributes substantially (probably $\gtrsim 50$\%) 
to the $R$-band flux in the central regions of several galaxies;
tidal tails do not necessarily accompany this star formation.
Many of the galaxies in our sample have bluer
centers than outskirts, suggesting that pre- or non-merger 
interactions may lead to evolution along the Hubble sequence.
These objects would appear blue and compact at higher redshifts;
the older, redder outskirts of the disks would be difficult to
detect.
Our data indicate that galaxies with larger separations on the sky
contain weaker, and probably older, bursts of star formation on average.
However, confirmation of these trends requires further
constraints on the colors of the older stellar populations and
on the reddening for individual galaxies.

\end{abstract}

\keywords{galaxies: evolution --- galaxies: fundamental parameters --- galaxies: kinematics and dynamics --- galaxies: interactions --- galaxies: structure}
}

\twocolumn[\head]

\section{Introduction}

\setcounter{footnote}{1}
\footnotetext{Hubble fellow.}

\citet{LT78} showed that galaxies in pairs exhibit a broader scatter
in the $U-B$/$B-V$ plane than ``field'' galaxies, demonstrating that
interactions trigger bursts of star formation.  Since then, numerous
studies, both of statistical samples and of individual interacting systems, 
have confirmed their results \citep[e.g.,][]{K84,m86,K87,js89,sw92,
Ke93,lk95a,lk95b,Ke96,dz97,BGK00}.
These studies have conclusively established the link between 
interactions and star formation.  However, many questions 
remain about the true role of
interactions in the evolutionary history of galaxies in the Universe.

Because interaction timescales are much shorter than a Hubble time, 
galaxies may participate in many interactions and/or mergers
during their lifetimes.  
The strengths and durations of typical triggered bursts of
star formation are therefore important cosmological parameters.
An understanding of interactions is a necessary basis for
including interactions and mergers
in models of galaxy formation \citep[e.g.,][]{SP99,Di99,Ka99a,Ka99b}.
Studies of local interactions also aid interpretation of the 
unusual morphologies and
structural parameters of intermediate redshift galaxies
\citep[e.g.,][]{K94,A96,BvZ01}.  Finally, a measure of the
total effects of interaction may yield constraints on the
causes of evolution along the Hubble sequence;
large amounts of gas funneled into the centers of the galaxies could
result in interaction-induced
forms of the processes in secular evolution \citep{PN90} 
that cause the
formation of ``exponential bulges'' \citep{AS94,C99}.

\citet[BGK hereafter]{BGK00} study 502 galaxies in pairs or
compact groups selected from the CfA2 redshift survey.  They find
a significant correlation between pair separation on the sky,
$\Delta D$, and H$\alpha$ equivalent width, \ew.  The correlation
also extends to the $\Delta V$/\ew\ plane, where $\Delta V$
is the pair separation in velocity. BGK argue that the 
$\Delta D$ -- \ew\ correlation
results from the aging of a continuing burst of star formation.
If their interpretation is correct, the correlation provides
a method of measuring the durations and initial 
mass functions (IMFs) of the bursts by comparing dynamical timescales with
star-formation timescales.  Here we investigate the
origin of this correlation by using $B$ and $R$ photometry to constrain
the old stellar population and the new burst of star formation
separately.

To measure the amount and current age
of the new burst of star formation, we explore a method of using the measured
colors and \ew\ to characterize
recent star formation superposed on a pre-existing stellar population.
We relate these quantities to
the orbital parameters of the pairs, thus exploring the origin of
the BGK correlation.
In Sec. 2 we describe the pair sample and the data.  Sec. 3 contains a
description of the ``two-population'' model we apply to characterize
the burst of star formation independent of the underlying older stellar
population.  We include a brief discussion of 
the results of applying the model
to the data.   We discuss the origin of the BGK $\Delta D$ -- \ew\ 
correlation in Sec. 4.  In Sec. 5 we explore the dependence of
burst strength and age on the R-band luminosities 
and rotation speeds of the galaxies.  We
describe the galaxies with strong bursts of central star formation
in Sec. 6 and conclude in Sec. 7.

\section{The Pair Sample}

We base our study on the sample of all 786 galaxies in pairs and compact groups
drawn from the original CfA2 redshift survey 
(m$_{\rm Zw} \leq 15.5$) with 
$\Delta D \leq 50$ h$^{-1}$ kpc, $\Delta V \leq 1000$~km/s
and $v \leq 2300$~km/s, where $\Delta V$ is the pair (or compact group
neighbor) velocity separation, and $v=cz$ is the apparent recession
velocity.  
BGK estimate that this ``full'' sample of 786 galaxies from the original 
CfA2 survey is 70\% complete 
with respect to all known galaxies in pairs 
over the region of interest in
the updated CfA2 redshift survey \citep{F99}. 
41 of the 502 galaxies with (new) spectra fail to
satisfy the original selection criteria.  We include these galaxies in
the analysis because most come close to satisfying the criteria; all
pairs in the sample satisfy $\Delta D < 77$~h$^{-1}$~kpc and 
$\Delta V < 1035$~km/s.

We divide the full sample into three overlapping sub-samples for which we
have different kinds of data: the dynamical sample ($\sim$140 galaxies, mostly
spirals and S0s, with
H$\alpha$ rotation curves), the spectroscopic sample (502 galaxies with
one-dimensional spectra; the same as the BGK sample) 
and the photometric sample (196 galaxies with calibrated
photometric images in B and R).

In this paper, we concentrate on the photometric properties of the galaxies in
the overlap of the spectroscopic and photometric samples
(190 galaxies comprising all or part of 
78 pairs and 13 compact groups; SP sample hereafter).  We have H$\alpha$ rotation
curves for 103 of these galaxies; 89 provide useful measurements
of $V_{\rm c}$, the velocity width.  BGK separate the sample into galaxies
residing in low- and high-density environments 
and concentrate on the low-density subset
to avoid contamination by the morphology-density relation and
gas stripping in clusters. Here our sub-samples are much smaller
and we are unable to divide the data
in this way.  However, we use direct measurements of V$_{\rm c}$ to
understand the galaxy mass-dependent effects.

In spite of the m$_{\rm Zw} \leq 15.5$ cutoff, the pair sample contains
numerous galaxies with $B$ magnitudes $> 15.5$.  These systems were often
recorded in the CGCG \citep{Z68}, and therefore the redshift catalog, 
with the Zwicky magnitude of the whole system instead of
separate magnitudes for each of the galaxies.  We find 33 (of 190) 
galaxies (17\%) with m$_{\rm B26} > 15.5$ in the SP sample, 
where m$_{\rm B26}$ is the $\mu_{\rm B} = 26$ isophotal magnitude.  
The sample is incomplete with respect to these galaxies.

We selected the systems in the dynamical sub-sample
based on the suitability of at least one galaxy for a rotation
curve measurement; therefore this sub-sample heavily favors
H$\alpha$-emitting galaxies and spirals less than $\sim$50$^\circ$ from
edge on.  
For the selection of the photometric and spectroscopic samples,
we focus on (1) galaxies in
the dynamical sample, and (2) galaxies known to have emission lines.
Therefore, the SP sample heavily favors emission-line galaxies.    
We measure the extent of the bias by comparing
the spectroscopic properties of the SP sample with 
those of the CfA2South pair sample, which is 100\% 
complete with respect to the set of all
pairs selected from the original CfA2 survey.
Although only 25\% of the complete CfA2South sample has
detectable \ew, 75\% of the (overlapping) 
SP sample has detectable \ew.

\begin{figure}[tbh]
\vspace{0cm}
\hspace{0cm}
\epsfxsize=\linewidth
\epsfbox{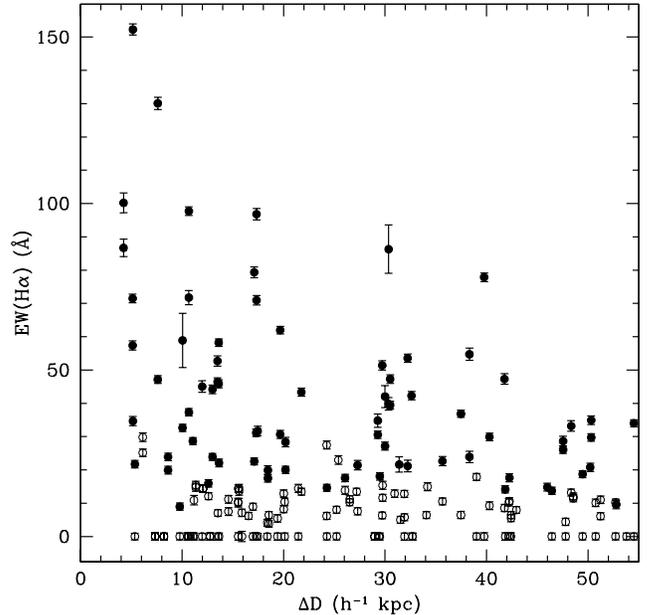}
\caption{The relationship between 
$\Delta D$ and \ew\ for the
190-galaxy SP sample (for comparison with BGK Fig.~2a).  The
trend is significant, with P$_{\rm SR} = 0.058$ in the full SP
sample and P$_{\rm SR} = 0.0017$ in the sample restricted
to galaxies with measured Balmer decrements ({\it filled circles}).
We exclude 2 galaxies with $\Delta D > 55$~h$^{-1}$~kpc from 
the plot. \label{fig:dd_ew}}
\end{figure}


The SP sample does not favor line-emitting galaxies
with particularly strong emission.
The distribution of \ew\ in the line-emitting fraction of the SP sample
is unbiased with respect to the \ew\ distribution for the complete 
CfA2South sample; a K-S test is unable to detect differences
in the \ew\ distribution of galaxies with $\mew \geq 10$~\AA\
in the (overlapping) SP sample (115 galaxies) and the 
CfA2South spectroscopic samples (98 galaxies, with 
44 in the SP sample), with P$_{\rm KS} = 0.55$ including all
galaxies, or
P$_{\rm KS} = 0.18$ excluding the overlapping galaxies.
We plot $\Delta D$ vs. \ew\ (uncorrected for reddening)
for the SP sample
in Fig.~\ref{fig:dd_ew}.

Differences are evident in the $\Delta D$ and redshift
distributions of the SP and complete
CfA2South samples.  For example,
the SP sample contains an excess of pairs with 
$\Delta D \sim 10$~h$^{-1}$~kpc and an excess of pairs
at lower redshift. These effects result from 
the large-scale structure differences between 
CfA2North and CfA2South \citep[see][]{F99}.  We use 
Spearman rank tests to evaluate the significance of the
correlations we describe below.  Therefore, the differences in
the redshift and $\Delta D$ 
distributions should only have a significant bearing
on our conclusions if we have different selection
effects in different ranges of $\Delta D$.
We are not aware of any such effects.  

\subsection{Observations and Data Reduction}

BGK and \citet{B01} describe the basic observations
and data reduction; we summarize these observations here
and we list the results in Table~\ref{tab:data}.

\begin{table*}
\tabletypesize{\scriptsize}
\tablenum{1}
\tablecolumns{12}
\caption{The SP Sample}
\begin{tabular}{rrrrrlrrrrrr}
\tableline
\tableline
\colhead{Coordinates} &
\colhead{} &
\colhead{$cz$} &
\colhead{$\Delta D$} &
\colhead{$\Delta V$} &
\colhead{} &
\colhead{} &
\colhead{EW(H$\alpha$)} &
\colhead{} &
\colhead{V$^{\rm c}_{2.2}$} &
\colhead{} \\
\colhead{(1950)} &
\colhead{N$_{\rm sys}$} &
\colhead{(km/s)} &
\colhead{(kpc/h)} &
\colhead{(km/s)} &
\colhead{M$_{\rm R}$} &
\colhead{$(B-R)_{\rm s}$} &
\colhead{(\AA)} &
\colhead{$\frac{{\rm H}\alpha}{{\rm H}\beta}$} &
\colhead{(km/s)} &
\colhead{Note} &
\colhead{Label}\\ 
\tableline
00 03 47.2 $+$17 09 05 & 2 &   5590 &   7.6 &    181 & -19.94 \tablenotemark{a}
\tablenotemark{b} &   0.75 &  130.1 &  4.07 &  & marg &   \\
00 03 47.5 $+$17 09 32 & 2 &   5770 &   7.6 &    181 & -20.14 \tablenotemark{a}
\tablenotemark{b} &   0.92 &   47.1 &  4.82 & 221.5 & norm &   \\
00 11 07.7 $+$47 52 24 & 2 &   5445 &  31.9 &     22 & -21.40 \tablenotemark{a}
\tablenotemark{c} &   1.39 &    0.0 &  --  &  &      &   \\
00 11 19.5 $+$47 52 47 & 2 &   5422 &  31.9 &     22 & -21.17 \tablenotemark{a}
\tablenotemark{c} &   1.51 &    5.8 &  --  & 372.6 & norm &   \\
00 17 52.0 $-$00 32 55 & 2 &   5222 &   9.8 &    423 & -20.84 \tablenotemark{a}
\tablenotemark{b} &   1.38 &    9.0 &  2.88 & 247.9 & marg &   \nl
\tableline
\end{tabular}
\tablenotetext{a} {Image was calibrated from a shorter exposure.}
\tablenotetext{b} {Galaxies overlapped.}
\tablenotetext{c} {Bright star in background, or smaller but significant star(s) on galaxy.}
\tablecomments{Galaxies in the SP sample:
(1) 1950 coordinates, (2) number of galaxies in system,
(3) redshift, (4) separation of pair on the sky
(or distance to closest galaxy in n-tuple), (5) velocity separation
of pair (or velocity difference to closest galaxy in velocity
within the n-tuple), (6) $R$ absolute magnitude, extrapolated
to the total magnitude and measured assuming
H$_{0} = 70$ km s$^{-1}$ Mpc$^{-1}$, (7) color of the portion
of the galaxy on the slit during the spectroscopic observation,
corrected for Galactic extinction and nebular lines,
(8) H$\alpha$ equivalent width, (9) Balmer decrement (corrected
for Balmer absorption), (10) corrected velocity width, and (11)
a description of rotation curve, where norm = ``normal'', marg =
``marginal'' (intermediate between normal and distorted),
and dist = ``distorted'' \citep[see text of][]{B01}, and
(12) label for Figs.~\ref{fig:strongburst} and
\ref{fig:turnaround}.  [The complete version of this table will appear
in the electronic edition of the Journal and is available
from the first author.]}
\label{tab:data}
\end{table*}

We observed the galaxies at the FLWO 48$^{\prime\prime}$ 
telescope on Mt. Hopkins
through either Johnson or Harris $B$ and Cousins or
Harris $R$ filters,
for total exposure times of 15 
minutes (usually spread over 3 images) and 5 minutes (usually
exposed all at once or spread over 2 images),
respectively.  
After calibrating the images to the \citet{L92}
standard-star system,
we subtract the sky background and measure magnitudes and surface 
brightness profiles with ellipse fitting on the $B$-band
images; we overlay $B$-band isophotes onto the $R$-band images.
\citet{B01} discuss the Galactic reddening correction and
several issues relating to galaxy distortion
and overlap.

\begin{figure}[tbh]
\vspace{0cm}
\hspace{0cm}
\epsfxsize=\linewidth
\epsfbox{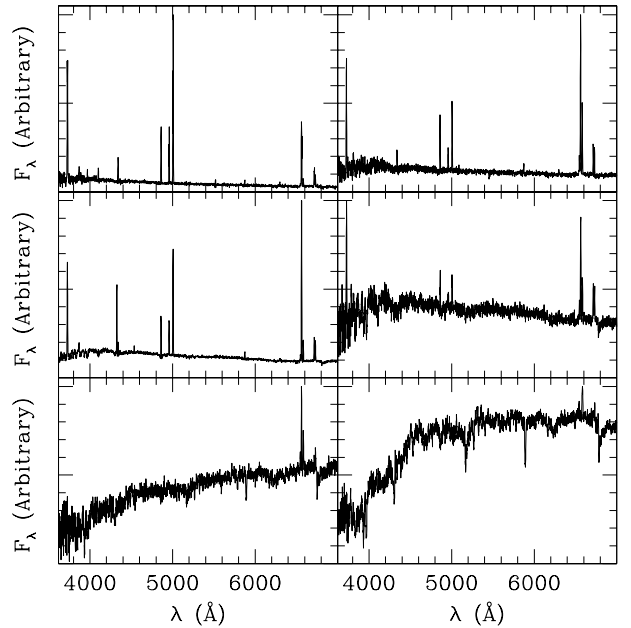}
\caption{Sample spectra illustrating the range of observed star-forming
properties of the interacting galaxies.  To show the quality of the
data, the spectra have not been smoothed. 
\label{fig:sample_spectra}}
\end{figure}

BGK contains a complete description of the spectroscopy and
data reduction.  We
observed the galaxies using the FAST spectrograph at the
1.5m Tillinghast telescope on Mt. Hopkins. Typical exposures
lasted $\sim$10~--~20 minutes.  We observed
the galaxies through a $3^{\prime\prime}$ 
slit using a grating with
300 lines/mm to disperse the light between 4000 and 7000~\AA.
The galaxies were often observed through clouds; thus,
the spectra are roughly calibrated only in a relative sense, using
observations of one or more spectrophotometric standards each night.
The photometry of the galaxies provides 
a more reliable indicator of the overall continuum shape of the
spectral energy distribution.  Fig.~\ref{fig:sample_spectra} shows examples
of the spectra, illustrating the observed
range of star-forming properties.

After flatfielding 
the data, we extract apertures which include
the brightest continuum light incident on the 
slit from the galaxy.  This
procedure results in a range of aperture lengths, from 
0.25~--~13.7~h$^{-1}$~kpc, with a median of only 2.4~h$^{-1}$~kpc.
Thus, strictly speaking, the spectra are not ``nuclear'' but 
``central'' spectra encompassing the central light distribution.
We compute equivalent widths that are the ratio of the flux in the line
to the measured continuum.  We base errors on photon statistics; repeat
measurements indicate that these errors are within a factor of $\sim$2 of the
true measurement errors.
We correct for Balmer absorption
by computing the maximum absorption equivalent width in H$\delta$ or 
H$\gamma$ and adding that equivalent width to both H$\alpha$ and
H$\beta$.

For the SP sample, we use the cleaned\footnotemark\ 48-inch images 
\citep[see][]{B01} to measure the $B$ and $R$ isophotal magnitudes, 
which we extrapolate to total magnitudes, and
$B-R$ color, $B_{\rm s}$, $R_{\rm s}$, and $(B - R)_{\rm s}$, respectively, 
of the light that accumulated 
through the slit during the spectroscopic observations.  
We compute these quantities
by adding the flux within a rectangular aperture measuring 
$3^{\prime\prime} \times L_{\rm ap}$ and aligned 
with the spectroscopic aperture (to within $\lesssim 1.5^{\circ}$),
for both the $B$ and $R$ clean images.
$3^{\prime\prime}$ is the width of the slit and $L_{\rm ap}$ is the (variable)
length of the extracted aperture.  We assume the slit center and
the center of the galaxy (the center of the smallest fit ellipse) are
coincident,
a close approximation to the actual position of the slit during
the spectroscopic observation.  We measure conservative 
errors in $(B - R)_{\rm s}$
due to misplacement of the slit center by displacing the
slit 2 pixels ($\sim1.2^{\prime\prime}$) in the $\pm$x and $\pm$y 
directions, for a subset of 30 galaxies.  
Changes in the color range from $\sim$-0.1 to
$\sim$+0.2 magnitudes, with $\sigma = 0.055$. 
When the spectrum consists of two or more exposures, 
usually with different values of 
$L_{\rm ap}$, we scale the fluxes in the different apertures by the exposure
times of the spectra; this procedure is only approximately correct in cases
where the atmospheric absorption differed during the
observations. 
\footnotetext{The ``cleaned'' images have been 
stripped of overlapping stars and, in some cases, overlapping 
companion galaxies; we usually replace the stripped regions with an 
elliptically symmetric model of the galaxy.}

Finally, we observed galaxies in the dynamical sample with the Blue Channel
Spectrograph at the Multiple Mirror
Telescope on Mt. Hopkins between November 1996 and February 1998.
For the majority of the galaxies, we used a 1$^{\prime\prime}$ slit
and a grating with with 1200 
lines/mm centered at $\sim$6500 \AA, 
and including roughly the wavelength range  5800~--~7200~\AA.

\citet{BKKG00} and \citet{B01} describe the data reduction in detail.
We use a cross-correlation technique that makes simultaneous
use of the major emission lines: [NII] ($\lambda$6548 and
$\lambda$6583), [SII] ($\lambda$6716 and $\lambda$6730), and
H$\alpha$; this technique is straightforward to implement and yields 
well-defined errors.  We use the method of \citet{C97}, 
developed for Tully-Fisher \citep[][TF hereafter]{TF77}
applications, to measure the velocity width, V$_{2.2}$, for galaxies
in the dynamical sample.  We fit each rotation curve with an
empirical fitting function \citep[see][Eqn.~2]{C97}.  \citet{B01}
describe the corrections we apply for inclination, slit misalignment,
and redshift, to arrive at V$^{\rm c}_{2.2}$, the corrected velocity width.

\section{Parameterizing the Central Bursts of Star Formation}

For lack of adequate constraints, full modeling of an
unresolved stellar population requires assumptions
about the star formation rate as a function of time.
Many studies parameterize the star formation rate with a 
single function, such as an exponential, and fit for the 
timescale, $\tau$ \citep[e.g.,][]{LT78,A96,BdJ00}; others
find the best-fit spectral energy distribution
from a set of empirical templates
\citep[e.g.,][]{TWS01}.  Because our goal is to
parameterize a recent ($\lesssim 1$ Gyr) burst 
of star formation superposed on an otherwise typical galaxy,
we must parameterize the ``old'' (pre-existing) and 
``new'' (interaction-related) populations separately.

\begin{table*}
\tablenum{2}
\tablecolumns{3}
\caption{Assumptions for Burst Model Contours}
\begin{center}
\begin{tabular}{ccc}
\tableline
\tableline
\colhead{} & \colhead{Assumption} &
\colhead{Other assumptions} \\
\colhead{Parameter} & \colhead{for Fig.~\protect \ref{fig:contours}} &
\colhead{from this paper or BGK} \\
\tableline
Pre-existing population   &  1.5 & 0.9 -- 2.0 \\
color, \BRo  &      & $1.5 + 0.75 \times [\log{\rm V^c_{2.2}}$-2.5]  \\
             &      & $1.5 + 1.50 \times  [\log{\rm V^c_{2.2}}$-2.5]  \\
             &      & \citet{PB96} bulges\\
             &      & NFGS distributions \\
\hline
Burst IMF                 & ``steep'' ($\alpha=3.3$) & ``Salpeter'' ($\alpha=2.35$) \\
                          &                                 & ``cutoff Salpeter'' ($\alpha=2.35$, Mass $< 30\ {\rm M}_{\sun}$) \\
\hline
Burst SFR(t)              &  Constant for 1 Gyr         &  Instantaneous \\
\hline
Burst metallicity         &  Solar                          &  0.4 Z$_{\sun}$, 2 Z$_{\sun}$ \\
\tableline
\tableline
\end{tabular}
\end{center}
\label{tab:contours}
\end{table*}

Several existing studies address this ``added-burst'' problem
\citep[e.g.,][]{LT78,K87,KTC94}.  
For all of these studies, there is a degeneracy associated with
separating the new burst from the pre-existing galaxy:  is the
galaxy blue because its star formation rate has been constant
or increasing with time throughout its history, or is it blue
because of a strong triggered burst?   Or, cast in terms of
the pair population as a whole, does the range of colors observed
in pairs result from the range of colors in the progenitor galaxies
or from triggered bursts of star formation?  One
approach involves beginning with the distribution of colors
vs. \ew\ observed in the field and exploring the effects of short
bursts of star formation \citep[e.g.,][]{K87}.
However, every point in the color/\ew\ plane
is reachable from a range of starting points.
In an approach similar to the models of
\citet{LT78}, we essentially fix assumptions about the pre-existing 
stellar populations and characterize bursts of
star formation in terms of burst age and ``strength'', 
testing the effects of modifying the initial assumptions.
We avoid assumptions about mass-to-light ratios by defining burst strength in 
terms of the fraction of the $R$-band flux originating
from the new burst.  In Secs.~3.2-4 we discuss the full range of possibilities;
we include a discussion of techniques for resolving the
degeneracy between young pre-existing stellar populations and 
triggered bursts.  We also discuss the effects of other properties,
like metallicity, on our results.  We demonstrate the effectiveness of
using only $B-R$ and \ew\ in Sec.~3.3.

\subsection{The Two-Population Model}

For simplicity, we begin by assuming that two distinct
stellar populations contribute to the flux incident
on the slit aperture during the spectroscopic observation: (1)
a new population, solely responsible
for any H$\alpha$ emission, which we denote by the
subscript n, with $B$ and $R$ fluxes, \fbn\ and \frn, respectively,
that turn on  at ${\rm t}=0$ and evolve as the burst ages,
and (2) a pre-existing, older population, denoted by the subscript $0$, 
with fluxes \fbo\ and \fro, which do not evolve significantly
on the burst timescales ($\lesssim 1$ Gyr).  We explore
the effects of H$\alpha$ emission from the pre-existing stellar
population in Sec.~3.2.
With these assumptions, the H$\alpha$ equivalent width
from spectroscopic measurements [\ewm],
slit $B$ and $R$ fluxes from calibrated photometry,
and starburst models for simple burst histories, we estimate both
the strengths and ages of the starbursts.

The old population increases the continuum around
H$\alpha$.  Therefore, the fractional strength of the (new) burst in R, 
$\msr \equiv \frac{\mfrn}{\mfro+\mfrn}$, governs
the relationship between \ewm\ and the H$\alpha$ equivalent width
of the burst alone, \ewb.  
The measured equivalent width is
\begin{equation}
\mewm = \msr \times \mkf \times \mewb,
\label{eqn:sr1}
\end{equation}
where \kf\ is a factor of order unity 
($0.9 \lesssim$ \linebreak
$\mkf \lesssim 1.4$) 
that corrects for the difference between the burst
strength in $R$ and the burst strength at the continuum near 
H$\alpha$.  \kf\ depends on the spectral energy distributions 
of the old stellar population (SED$_{0}$) and the burst
(SED$_{\rm b}$).
We estimate \kf\ at the appropriate redshift for a given old stellar population and burst
by using \citet{BC96} 
Salpeter $\tau$-model spectral energy distributions for the old population,
matched in $B-R$ to the nearest $\tau$ model at 14 Gyr from formation
(with $\tau$ equal to an integral number of Gyr in most cases).  We
interpolate the \citet{L99} burst 
model spectral energy distributions for the burst SED computed 
at a limited set of burst ages.  We use the $R$ filter function of 
\citet{B90} and the $B$ filter function of \citet{BK78}.
In effect, we parameterize the old stellar population completely
via \BRo\, which is an input parameter, and \kf, which we compute from the
\citet{BC96} 
models based on the input \BRo.  

The metallicity of the old stellar population has little effect on \kf. 
For a fixed $\mBRo = 1.5$, we estimate the effects of metallicity by
computing \kf\ over the full range of strengths from 0 to 1 and
the full range of ages, IMFs, metallicities, 
and star foramtion histories of the L99 models.  The maximum difference in \kf\ 
introduced by assuming a non-solar metallicity (of either 0.4 Z$_{\sun}$ 
or 2.5 Z$_{\sun}$) is $< 1\%$.  Hereafter, we use Z$={\rm Z}_{\sun}$ 
to compute \kf.

The strength of the new burst also depends on the colors
of the old and new stellar populations.  
The composite (measured) $R$-to-$B$ flux 
ratio is
\begin{equation}
\mcs = \frac{\mco}{1+\msr \frac{[ \mco - \mcn ]}{\mcn}},
\label{eqn:sr2}
\end{equation}
where \co\ and \cn\ are the $R$-to-$B$ flux ratios
of the light incident on the slit from
the old and new populations respectively.
The $R$-to-$B$ flux ratios are related to the colors:
$\mco = \frac{\mfro}{\mfbo}= 10^{\mBRo/2.5}$, 
$\mcs = \left(\frac{\mfro+\mfrn}{\mfbo+\mfbn}\right) = 10^{\mBR/2.5}$, 
and $\mcn = \frac{\mfrn}{\mfbn} = 10^{\mBRn/2.5}$.

The ``strength'' parameter in Eqns.~\ref{eqn:sr1} and \ref{eqn:sr2}, 
\strr, is time-dependent. It represents the {\it current} fraction
of the central $R$-band flux originating from the new burst; 
this fraction increases as the burst builds luminosity and 
decreases as the burst fades.  We also consider
a time-independent parameter, \stro, the flux at an
age of 100 Myr  
[$\mso \equiv {\rm s}_{\rm R}({\rm 100 Myr})$]. 
Then at any other time,
\begin{equation}
\msr = 
\frac{\frac{\mso}{1-\mso} 10^{0.4[{\rm M_R(100\ Myr)-M_R(t)}]}}
{1+\frac{\mso}{1-\mso} 10^{0.4[{\rm M_R(100\ Myr)-M_R(t)}]}},
\label{eqn:sr3}
\end{equation}
where ${\rm M_R(t)}$ is the $R$-band absolute magnitude of the burst
as a function of time.

\subsubsection{Strengths and Ages from the Leitherer et al. Models}
 
Starburst models predict the
burst color, \BRn, and the burst H$\alpha$ equivalent width,
\ewb, as a function of time.  
We use the \citet[][L99 hereafter]{L99} starburst models
for burst parameters.
L99 model a starburst by ``forming'' stars according to an
IMF with $\frac{\rm dN}{\rm dM} = C{\rm M}^{-\alpha}$,
where M is the stellar mass, $\alpha$ is the slope of the IMF,
and $C$ is a constant. 
With models from the literature or their own models
for stellar evolution, stellar atmospheres, and nebular emission,
they compute EW(H$\alpha$)
as a function of time for both the ``instantaneous'' starburst case, 
with a timescale $\tau \lesssim 10^6$ years, and the continuous case, 
with star formation over $10^9$ years.  
 
The L99 models have a range of metallicities (0.05 Z$_{\sun}$, 
0.2 Z$_{\sun}$, 0.4 Z$_{\sun}$, Z$_{\sun}$, 2Z$_{\sun}$), 
and a range of IMFs, including an approximate 
\citet{S55} IMF, with $\alpha=2.35$, 
from 1~--~100~M$_{\sun}$, a Salpeter IMF with a 
30~M$_{\sun}$ high-mass cutoff, and a steeper function
with $\alpha=3.3$ from 1~--~100~M$_{\sun}$ patterned 
after the high-mass segment of the 
\citet{MS79} IMF.  Hereafter, we refer to these 
IMFs as Salpeter, cutoff Salpeter, and steep, respectively.

\begin{figure}[tbh]
\plotone{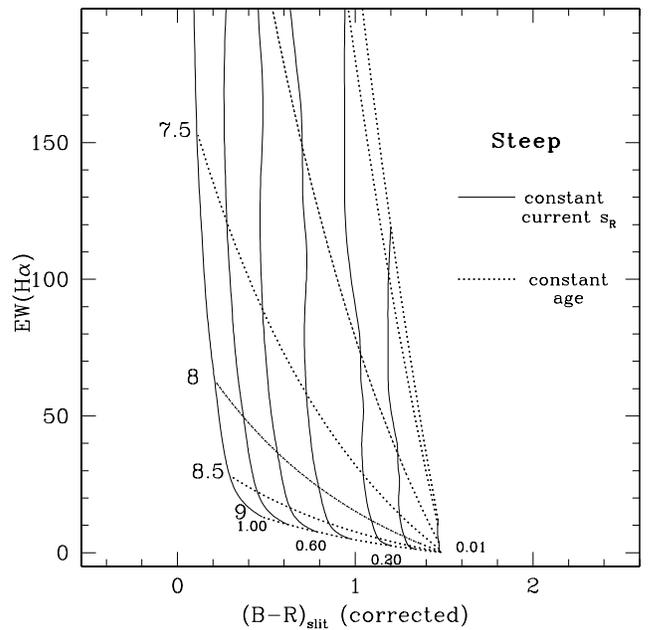}
\caption{Contours of constant current burst strength and age
in the plane of the observed parameters,
for a continuous burst of star formation (see Table~\ref{tab:contours}).
The dotted lines are lines of constant burst age, t, in
log(years); 
solid lines are contours of constant {\it current} strength, \strr, defined 
as the fraction of $R$-band flux from the new burst in the spectroscopic 
aperture. }
\label{fig:contours} 
\end{figure}

\begin{figure*}[tbh]
\epsfxsize=3.4in
\epsfbox{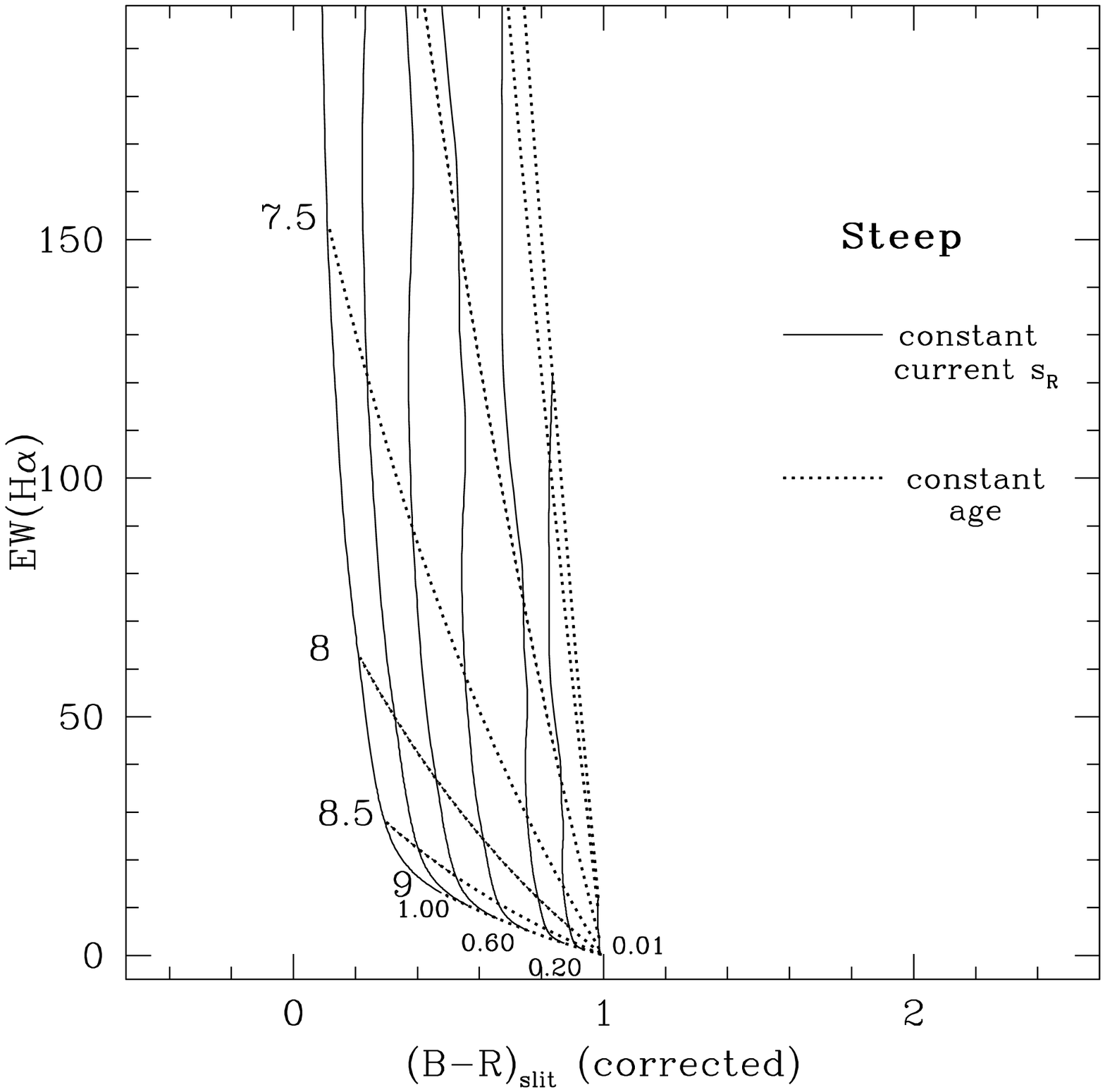}
\epsfxsize=3.4in
\epsfbox{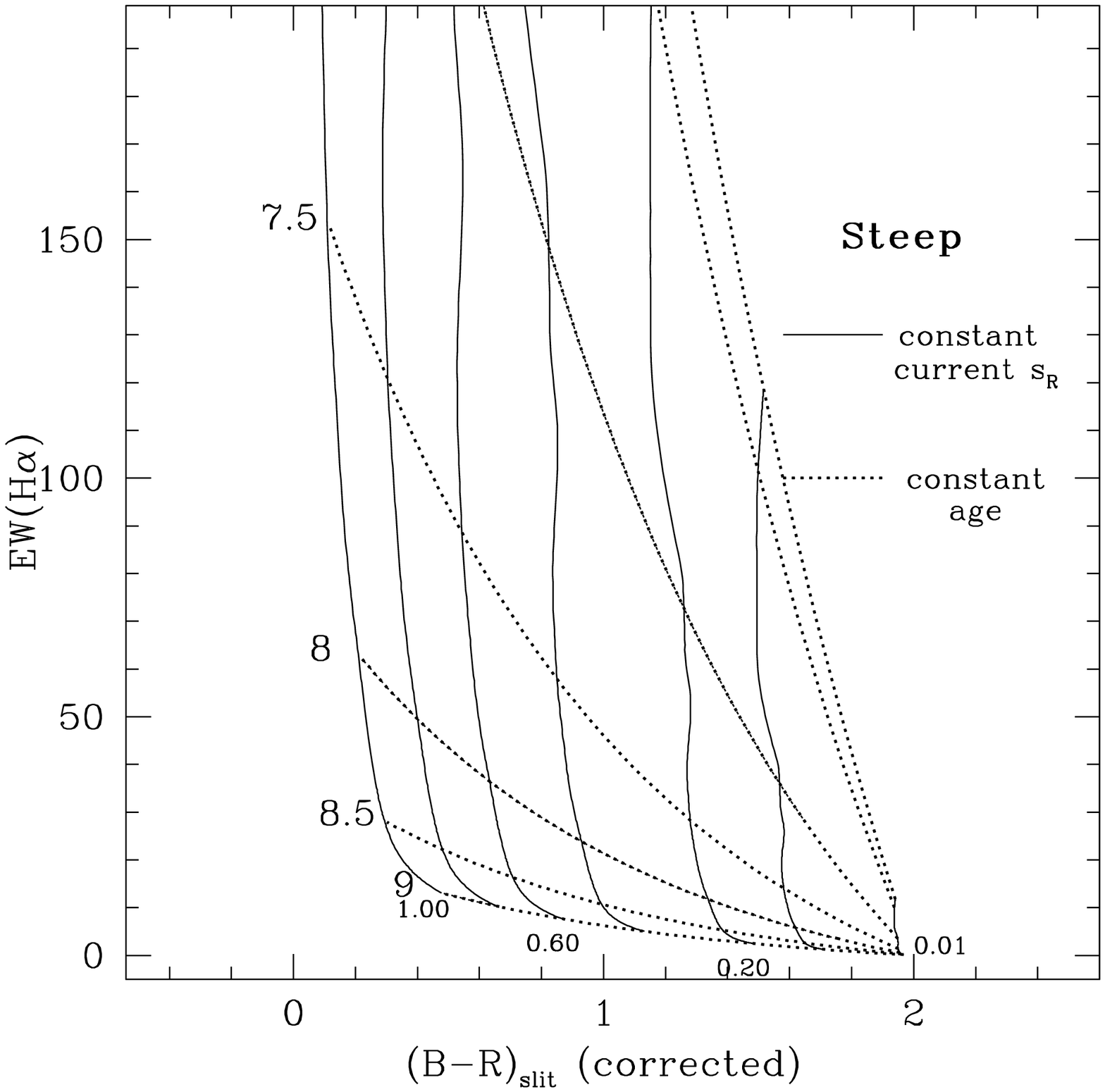}
\caption{Same as Fig.~\ref{fig:contours}, but with
$\mBRo = 1.0$ ({\it left}) and $\mBRo = 2.0$ ({\it right}).  Lines of
constant current strength, \strr, remain roughly vertical, but the
lines corresponding to weaker bursts are displaced depending on \BRo.
The blue edge of the contours is independent of \BRo; the IMF and 
the starburst models determine the location of a population that
is 100\%  triggered burst.}
\label{fig:contshift}
\end{figure*}

For an assumed pre-existing stellar population, 
the starburst models and Eqns.~\ref{eqn:sr1}~--~\ref{eqn:sr3} 
provide a method of estimating burst strengths, \strr\ or \stro, and 
ages, t,  from the measured quantities [\BRs\ and \ewm].  
Fig.~\ref{fig:contours} shows contours of 
constant current burst strength, \strr,
and age, t, in the (measured) \ew/\BRs\ plane for a 
continuous burst of star formation with the ``steep'' IMF
and $\mBRo = 1.5$.  Hereafter, we assume $cz = 5000$~km~s$^{-1}$ to compute
\kf, except for specific galaxies, for which we use the measured redshift.
Table~\ref{tab:contours} summarizes the other assumptions we make to construct
these contours.  For comparison, 
Fig.~\ref{fig:contshift} shows the contours of
constant (current) burst strength assuming old population colors of
$\mBRo = 1.0$ and $\mBRo = 2.0$.  

Lines of constant t$_{\rm burst}$ are diagonal in 
Figs.~\ref{fig:contours} and \ref{fig:contshift}, while
the contours of constant \strr\ are nearly
vertical.  Thus, for continuous bursts of star formation with 
steep IMFs superposed on fixed existing
stellar populations, the burst age, t$_{\rm burst}$, primarily
affects the measured \ew.  In contrast,
the strength of the burst in $R$, \strr, affects
both the measured \ew\ and the measured color,
\BRs.

\begin{figure}
\plotone{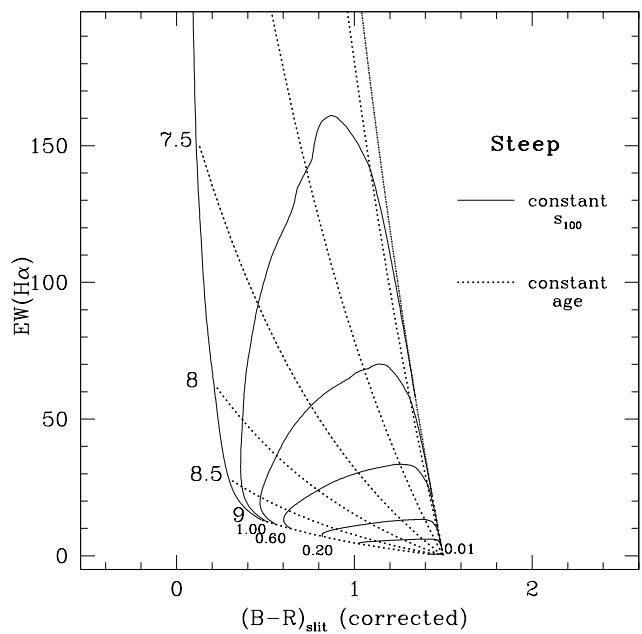}
\caption{Same as Fig.~\ref{fig:contours}, except that 
solid lines are contours of constant strength for
the burst at 100 Myr.  Thus, constant \stro\ lines track the evolution
of a continuing burst of star formation.  Unlike \strr, the measurement of
\stro\ depends heavily on
the assumption the burst star formation rate is constant.}
\label{fig:contours_s100} 
\end{figure}

Contours of constant \stro\ in 
Fig.~\ref{fig:contours_s100} are the evolutionary tracks of 
constant bursts of star formation within this model.  
\stro\ is the burst strength in $R$ that 
would be measured at 100 Myr, assuming that the burst 
continues (or has continued) at a constant rate.  
\stro\ depends heavily the assumption of a constant star formation 
rate.  Mass-fraction 
measures \citep[e.g.,][]{LT78,K87} require similar assumptions;
they incorporate assumptions about the relative 
mass-to-light ratios of the burst and the underlying old stellar
population.  

The contours in Fig.~\ref{fig:contours_s100} fold over and
are no longer vertical.  A moderately strong, young burst
continuing at its current rate dominates the 
flux of the galaxy at 100 Myr.
For example, within this model, if a continuous burst superposed on an
older population [$\mBRo = 1.5$] has $\mewm = 120$~\AA\
and $\mBRs = 0.6$, its current age (Fig.~\ref{fig:contours})
is $\sim$16 Myr and $\sim$50\% of the $R$-band flux currently 
originates from this burst.  If this burst
continues at its present rate, at 100 Myr it will
comprise $\sim$80\% of the $R$-band flux in the aperture
(Fig.~\ref{fig:contours_s100}).  
If the burst actually ceases at 316 Myr, it will comprise much
less than the original 
50\% of the R-band flux in the aperture at 100 Myr.

\subsection{Measuring Strengths and Ages from the Data}

The discussion in Sec.~3.1.1 omits several corrections necessary to apply
the models to the data.  
Table~\ref{tab:corrections} summarizes the corrections
we discuss in this section.  They include 
(1) the contribution from line emission to \BRs, which we estimate
using the measured equivalent widths of the nebular lines along with the
continuum from the model spectral energy distributions.  
We apply the nebular emission
(and absorption) corrections to the color data in the slit regions only (not to
the total magnitudes), and include the corrections in 
Figs.~\ref{fig:contours}~--~\ref{fig:contours_s100}; the
corrections range from -0.079 to 0.003 
with an average of -0.017,
(2) extinction corrections to the measured colors and/or \ewm, 
and (3) ongoing star formation in
the pre-existing stellar population.  Finally, we discuss our
choices for the color of the pre-existing stellar population,
\BRo.

\begin{table}[tbh]
\tablenum{3}
\begin{center}
\tablecolumns{3}
\caption{Corrections to the Data}
\begin{tabular}{ccc}
\tableline
\tableline
\colhead{} & \colhead{Correction(s) explored} \\
\colhead{Correction} & \colhead{in this paper} \\
\tableline
\tableline
{\small nebular line contamination in color}  & 
{\small corrected data using measured }\\
     & {\small equivalent widths, model SEDs}\\
\tableline
{\small Internal reddening}  
&  {\small ``traditional''\tablenotemark{a}} \\
                    & {\small none} \\
                    & {\small ``Calzetti''\tablenotemark{b}} \\
\tableline
{\small ongoing star formation in}  
& {\small no correction applied} \\
{\small pre-existing stellar population} &  \\ 
\tableline
\tableline
\end{tabular}
\tablenotetext{a}{See Sec.~3.1.}
\tablenotetext{b}{\citet{CKS94}; see Sec.~3.1.}
\label{tab:corrections}
\end{center}
\end{table}

\subsubsection{Corrections for Extinction}

Corrections for internal extinction substantially
modify the inferred ages and strengths of the bursts of 
star formation.  
Because it is independent of the
star formation history of the galaxy,
the Balmer decrement\footnotemark\ is generally
the best extinction measure available in our dataset.
However, several circumstances
affect our ability to measure the Balmer decrement and to
derive the appropriate extinction correction for the
galaxy.  
The presence of an intermediate-age ($\sim$Gyr) population
necessitates a correction for Balmer absorption under the
emission lines (see Sec.~2.1 and 
BGK for a description of our correction). 
In addition,  \citet{CKS94} show that 
the extinction of the continuum is only half 
of that appropriate for line-emitting regions.
\footnotetext{The Balmer decrement is the flux ratio
H$\alpha$/H$\beta$, where both fluxes have been corrected
for stellar absorption with the prescription described 
in \citet{BGK00}.}

\begin{figure}
\plotone{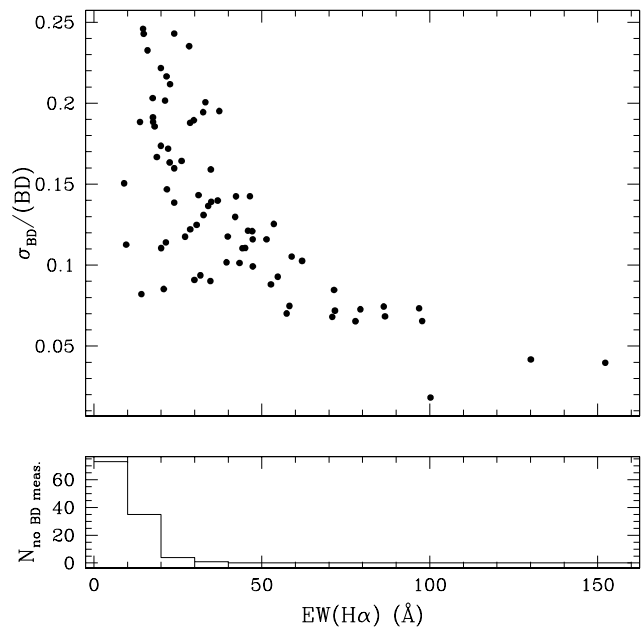}
\caption{({\it Top}) Fractional error in the Balmer decrement as a 
function of \ewm.  Our cutoff is an error of 25\% of the measured
value.  ({\it Bottom}) Histogram of the galaxies with $> 25$\% errors.
The set of galaxies with measured Balmer decrements is incomplete
for $\mewm \lesssim 20 - 30$~\AA.}
\label{fig:BalmerDec}
\end{figure}

The difficulty of measuring the Balmer decrement in galaxies with
insufficient star formation introduces selection effects into
the extinction-corrected sample.
The top panel of Fig.~\ref{fig:BalmerDec} shows the fractional
error in the Balmer decrement
as a function of EW(H$\alpha$) for the 78 galaxies with accurate
measurements (better than 25\%); 
112/190 galaxies yield larger errors in 
${\rm H}\alpha/{\rm H}\beta$ or exhibit no significant
H$\beta$ emission.
The bottom panel shows a histogram of \ewm\ for the galaxies
with no ${\rm H}\alpha/{\rm H}\beta$ measurements; 
incompleteness in the sample of galaxies with accurate Balmer 
decrement measurements begins at $\mewm \sim 20$~\AA.
In the analysis below, 
we primarily use the 78 galaxies with accurate measures;
we replace values of
${\rm H}\alpha/{\rm H}\beta < 2.85$ with 2.85 (or 2.88 as necessary).

We explore the possible effects of dust on our data by
considering three different dust corrections: no correction,
a ``traditional'' correction with no differences in extinction 
between the continuum and the line-emitting regions, and the extinction
correction of \citet[][``Calzetti'' correction, hereafter]{CKS94}.

The ``traditional'' correction involves no adjustments to \linebreak 
\ewm. We correct \BRs\ for reddening assuming 
\begin{displaymath}
{\rm E}(B-R) = 1.78{\rm E}(B-V) = 1.536{\rm E}(\beta - \alpha) 
\end{displaymath}
\begin{displaymath}
\equiv 1.536 \times 2.5 \log\left[\frac{({\rm H}\alpha/{\rm H}\beta)}{2.85}\right]
\end{displaymath}
\citep{Z90,MM72,V95}.
Here, E$(B-R)$, E$(B-V)$, and E$(\beta - \alpha)$ are
the $B$-to-$R$, $B$-to-$V$, and H$\beta$-to-H$\alpha$
reddening, respectively, and ${\rm H}\alpha/{\rm H}\beta$ is
the flux ratio of H$\alpha$ to H$\beta$, corrected
for Balmer absorption.  The error in the measured 
H$\alpha$/H$\beta$ includes a small addition for 
uncertainty in the Balmer absorption correction.

The ``Calzetti'' correction is appropriate if dust is clumped around regions
of recent star formation \citep[e.g.,][]{CF00}; the prescription
requires separate corrections for
the continuum and line emission \citep{CKS94}:
\begin{displaymath}
{\rm E}(B-R) = 0.8170 \ln{\left[\frac{\rm H\alpha/H\beta}{2.88}\right]}
\end{displaymath}
and
\begin{displaymath}
{\rm EW(H{\alpha})^{c}} = \mewm \times 
\left[ \frac{\rm H\alpha/H\beta}{2.88} \right]^{0.9005}.
\end{displaymath}

\subsubsection{Star Formation in the Pre-existing Stellar Population}

In this section, we discuss our choice not to correct for ongoing star 
formation in the pre-existing stellar population.  
The appropriate corrections depend on the intended interpretation 
of the results.  The fundamental question is whether the new burst of star 
formation is physically independent of the ongoing star formation in
the progenitor galaxy.  
If the pre-existing stellar population and the recently-formed
burst population are truly independent of one another, 
ongoing star formation in the 
pre-existing (``old'') population necessitates corrections
to the data and models.
Non-zero corrections to H$\alpha$ result in weaker burst strengths 
and older burst ages.  We argue, however, that once an interaction disrupts
a galaxy, star formation is not independent in the pre-existing and burst populations. 
The process that triggers the new, added burst of star formation likely disrupts the
``quiescent'' star formation in the galaxy.  The triggered burst may 
consume the gas that was forming stars in the absence of an interaction,  
invalidating the assumption of physically
independent pre-existing and burst populations.

\begin{figure}[tbh]
\plotone{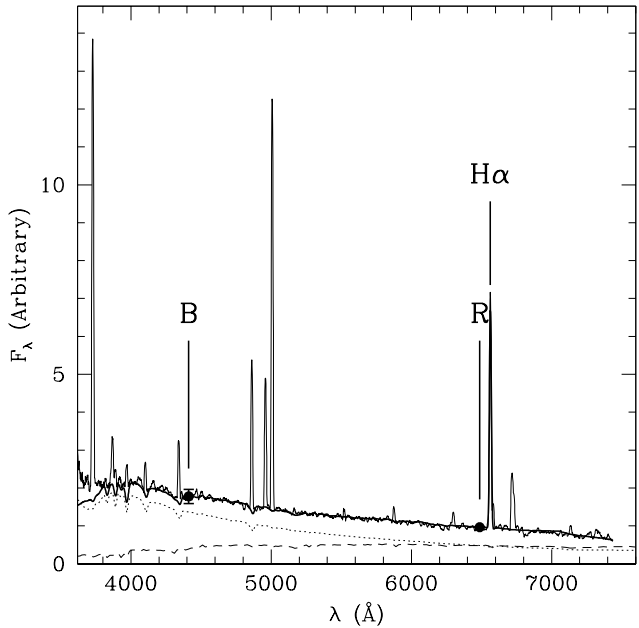}
\caption{An example of the model spectrum and the
galaxy spectral energy distribution for the ``clean'' case of NGC~2719A,
which has a Balmer decrement consistent with no reddening.  The thick
black line is the spectral energy distribution of the
model, fit to the $B-R$ color and \ew\ of the center of the galaxy only.
The model is the sum of the relevant L99 model, a 21-Gyr-old burst with
a ``steep'' IMF, and the \citet{BC96} 
model with a Salpeter IMF  
and an exponential star formation timescale of
$\tau = 3$~Gyr, observed 14 Gyr after formation.  The
other nebular emission lines are not modeled.  We plot broad band 
fluxes as filled circles; $R$ is normalized to the spectrum and 
$B$ is plotted at its effective wavelength.  The data and models
have both been smoothed by 10 bins. }
\label{fig:nogal_029_modelfit}
\end{figure}

Without physical distinctions between the pre-existing and
triggered populations, the amount of triggered star formation
is an ill-defined quantity.  
Two better-defined, distinct quantities are:
(1) the statistical increase in central star formation 
due to major interactions, and (2) the total star formation 
ongoing in the centers of interacting galaxies (including
star formation that would have occurred in the absence of
a major interaction).  Measuring these quantities
requires different approaches to the correction for 
ongoing star formation in the pre-existing stellar population.
The first  requires a statistical correction for 
quiescent star formation; the second question requires no correction, but a
judicious choice for \BRo\ that allows for the possibility of
star formation just prior to the interaction.

Regardless of the interpretation, 
a moderate amount of ongoing star formation
in the pre-existing stellar population has few 
qualitative effects.  The primary effect is to place galaxies with
no interaction-triggered star formation onto the apparent
burst contours (as in Fig.~\ref{fig:contours}).
These galaxies occupy a locus 
across the lower portion of the \BRs/\ewm\ plane defined 
by ``normal'' galaxies.  For these galaxies, the model analysis is 
not applicable.  Thus, some of the galaxies that appear, in this
model, to have weak bursts of star formation may actually be 
quiescent late-type galaxies; our
analysis is more reliable for galaxies with larger \ewm.

In the sections that follow, we focus on the
approach of characterizing the physical conditions in
galaxies with triggered star formation.  We apply no corrections 
for ongoing star-formation directly to the data, but explore
the effects of different choices for \BRo.

\subsection{The Scope of the Model}

\begin{figure*}[tbh]
\epsfxsize=\linewidth
\epsfbox{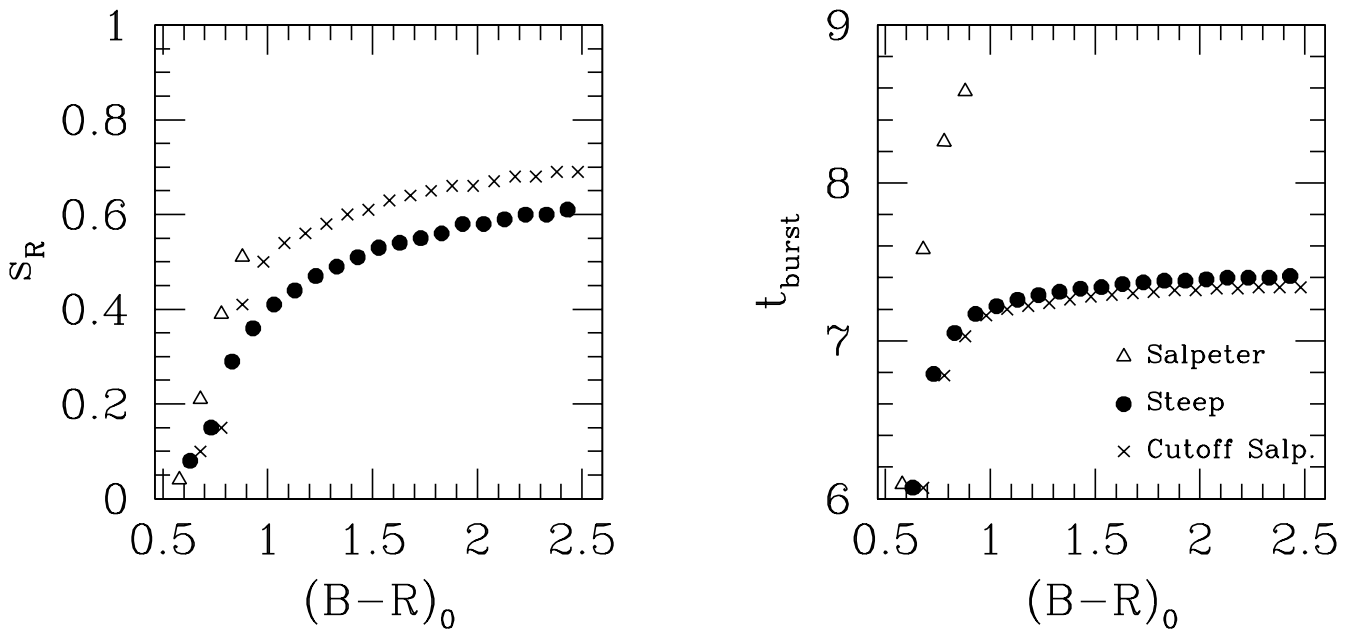}
\caption{Solutions for burst strength, \strr\ ({\it left}), and age, 
t$_{\rm burst}$ ({\it right}) for NGC~2719A as a function of \BRo, 
for the three different IMFs, assuming solar metallicity.
NGC~2719A has $\mew = 100$~\AA\ and $\mBRs = 0.56$ with no evidence for
extinction in the optical spectrum.}
\label{fig:solutions}
\end{figure*}

To characterize the star-forming properties of our galaxies, our 
approach is to fit simple star-formation models to the observed 
$B-R$ colors and \ew\ in the centers of the galaxies, using spectral
synthesis models from the literature \citep{L99,BC96}.  
These models predict the entire spectrum of the galaxy, 
with the exception of
most of the nebular line strengths; the agreement between the
predicted and observed spectra serves as a consistency
check of our models.
Fig.~\ref{fig:nogal_029_modelfit} shows the ``clean'' 
case of NGC 2719A, with a measured Balmer decrement 
that yields no evidence for reddening in the spectrum. 
The thin solid line is the data; the thick solid line 
is sum of the model spectra from 
\citet{BC96} 
and
L99, respectively; we also include the H$\alpha$ line, with the
appropriate measured \ew.
For the old stellar population ({\it dashed line}), we use a 
\citet{bc96}
$\tau=3$ model with a Salpeter IMF at 14 Gyr, corresponding to 
$\mBRo=1.46$ [close to our fiducial $\mBRo = 1.5$].
The measured $\mBRs = 0.56$ and $\mew = 100$~\AA\ yield a burst 
strength and age of 0.51 and 21.4 Myr, respectively, for this 
\BRo\ and the ``steep'' IMF, with solar metallicity.  
Hence, the ``new burst'' spectrum ({\it dotted line}) 
is the L99 model with continuous star formation and ``steep'' 
IMF after 21 Myr.  The thick black line is the sum of these models,
combined according to the derived value of \strr.
Filled circles represent measured $B-R$ color, plotted at the effective
wavelength of each band with both filter passbands from \citet{B90}; 
we normalize $R$ and plot $B$.
By design, the broad-band color agrees perfectly with the model.

Apart from the nebular emission lines, which are not predicted by the models,
the spectral shape agrees well with the model, even though it
is not used as an input.  This agreement results from the agreement
between the rough calibration of the spectrum and the more reliable
broad band colors, which are generally within $\sim$0.2 magnitudes of
one another for the galaxies with no reddening.  The exception is the very 
blue end, below $\sim$4000~\AA, where the calibration of the spectrum is 
highly uncertain.  The good agreement between the predicted spectrum
and the observed spectrum justifies our use of only the broad band colors
and \ew\ {\it a posteriori}: the spectra do not contain any more 
of the information predicted by the model than we are already using.
In particular, the model spectra corresponding to other \BRo\ values
[and adjusted according to the new derived \strr\ and t$_{\rm burst}$]
appear to agree similarly well, within the limits of the expected
accuracy of the calibration of the spectrum, 
indicating that the optical spectra contain 
no additional reliable information about the color of the pre-existing 
stellar population.

Broadly speaking, the spectral energy distribution of a galaxy with 
a triggered burst of star formation is determined by a large number
of factors, including the metallicities, star formation histories,
and IMFs of both the old and new stars, and the reddening appropriate for
each stellar population.  As indicated above, the current dataset does
not provide separate information about all of these quantities.
Our approach is to identify the most significant factors and focus on their 
effects.  As we describe in Sec.~3.1,
for the fiducial $\mBRo=1.5$, the metallicity of the old stellar 
population only affects \kf\ by
$< 1$\%.  In Sec.~3.4, we characterize the effects of the 
metallicity of the burst and show that they are
dwarfed by uncertainties in the burst IMF and old stellar population color.
Hence, we focus on the most important factors, the 
star formation histories, burst IMF, and reddening
corrections.  Our approach, as described above, is to parameterize
the burst star formation history in terms of an age, t$_{\rm burst}$,
and a current $R$-band strength, \strr, and compute the results for
a set of different assumptions about the burst IMF, the reddening,
and the old stellar population star formation history and IMF [as
characterized together by \BRo].  Our argument is statistical in nature;
we attempt to account for the relationship between the ensemble properties
of the sample and the predictions of the models.

\subsection{Results}

The measured \strr\ and t$_{\rm burst}$ 
depend relatively strongly on the input color of the
old stellar population, \BRo. 
Fig.~\ref{fig:solutions} shows this dependence for NGC~2719A, the
galaxy depicted in Fig.~\ref{fig:nogal_029_modelfit}
with a strong burst of star formation and no apparent reddening.
The realistic possibilities for 
\BRo\ do not include the entire range in the figure (from 0.5~--~2.5).
For galaxies of types S0 to Sbc,
\citet{PB96} find bulge colors in the
range of $1.15 \leq B-R \leq 1.8$;
centers of later-type spirals
and irregulars are as blue as $\sim$0.8 
for ``field'' galaxies \citep{BdJ00}.  

\begin{figure*}[tbh]
\epsfxsize=\linewidth
\epsfbox{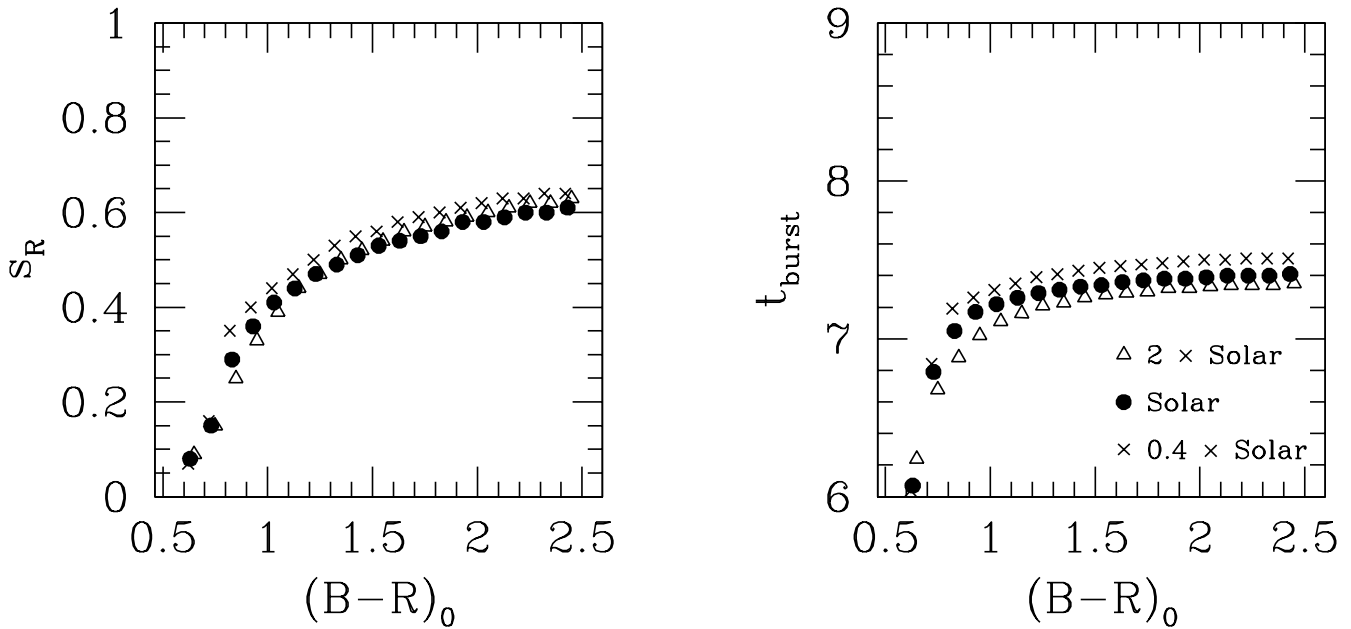}
\caption{Solutions for burst strength, \strr\ ({\it left}), and age,
t$_{\rm burst}$ ({\it right}) for NGC~2719A, as a function of \BRo,
for models with three different burst metallicities and a ``steep'' IMF.
Comparison with Fig.~\ref{fig:solutions} indicates that the uncertainties
in the assumptions about IMF and old stellar population color
dominate over the effects of assumptions about metallicity.}
\label{fig:nogal_029_met}
\end{figure*}

The uncertainties in the assumptions about IMF and \BRo\
dwarf the uncertainties introduced by assuming solar metallicity for the old
and new bursts.  As discussed in Secs.~3.1 and 3.3, uncertainties introduced
by the metallicity of the old stellar population are negligible.
Fig.~\ref{fig:nogal_029_met} illustrates the typical
effects of our assumption of solar metallicity for the burst; the effects 
are small compared with the other uncertainties.
To verify that the observed range of metallicities in our sample is
limited, we use the ``combined'' metallicity estimator of 
\citet{KD02}; 99/190 of the
SP sample spectra exhibit strong enough emission 
lines to make this measurement.  
For a solar abundance of $\log{(O/H)}+12 = 8.69$ \citep{APLA02}, the 
metallicities range from 0.42 Z$_{\sun}$ to 4.01 Z$_{\sun}$ with a 
median of 1.69 Z$_{\sun}$.  For the galaxies with measured Balmer
decrements, 4 have metallicities closer to the L99 model 
with Z$ = 0.4 {\rm Z}_{\sun}$ and 47 have metallicities
closer to the model with Z$ = 2 {\rm Z}_{\sun}$.  We test the differences
between using the solar metallicity model and the model closer in
metallicity. Assuming $\mBRo = 1.5$, 
the {\it maximum} differences in \strr\ and t$_{\rm burst}$ are 
0.06 and 15.9 Myr, respectively.
Thus, we assume solar metallcity hereafter, which has
no qualitative effects on our conclusions.

Our data are insufficient to distinguish among
different values of \BRo.  However, optical-to-near-infrared
colors will aid in the process.  
For example, a galaxy like NGC~2719A, with measured $\mew = 100$~\AA\ and
$\mBRs = 0.56$ has equally valid solutions for steep-IMF bursts of
(among others) $\mstr = 0.54$ and t$_{\rm burst}= 23$~Myr if
$\mBRo = 1.64$ and $\mstr = 0.3$, t$_{\rm burst} = 12$~Myr if
$\mBRo = 0.86$.  With a Salpeter-IMF burst on the same pre-existing
blue galaxy, the solution differs even more significantly in age, with 
t$_{\rm burst}=329$~Myr and $\mstr = 0.48$.
Using \citet{BC96} 
Salpeter IMF
$\tau$ models for the pre-existing galaxies, 
with $\tau = 1$~Gyr and $\tau = 100$~Gyr (i.e., a constant star formation
rate), respectively,
we predict differences in the burst+pre-existing galaxy
$V-K$ colors of 0.17 magnitudes
between the two solutions with steep IMFs, and 0.29 magnitudes between
the steep IMF solution on the red pre-existing population
and the Salpeter-IMF solution on the blue pre-existing population,
with the redder $V-H$ colors for the $\tau = 1$~Gyr Salpeter
pre-existing population model and the steep-IMF burst.
Thus, near-infrared colors help to distinguish among star formation 
histories and IMFs.

\begin{figure*}[tbh]
\epsfxsize=\linewidth
\epsfbox{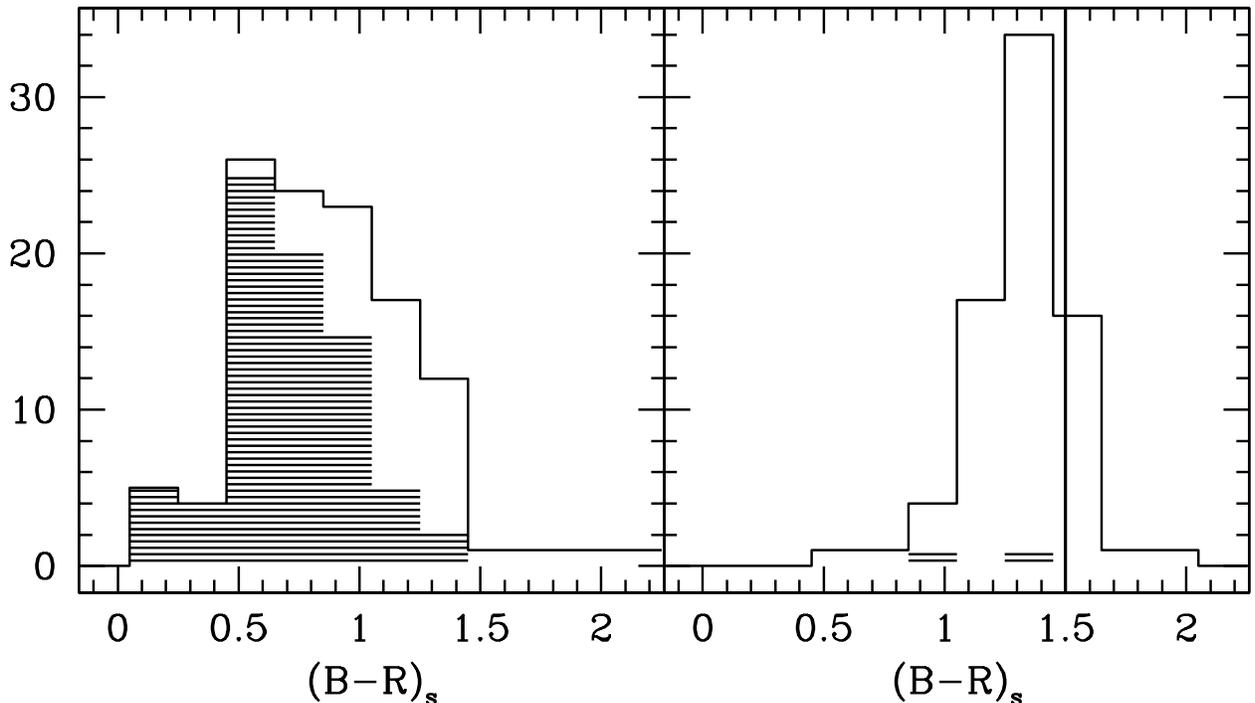}
\caption{$(B-R)_{\rm s}$, corrected for reddening and nebular lines with the
``Calzetti'' prescription, for: 
(a) line-emitting galaxies, with EW(H$\alpha$) $> 10$~\AA, and
(b) galaxies with EW(H$\alpha$) $< 10$~\AA.  The vertical line 
shows our choice of $\mBRo=1.5$.}
\label{fig:oldpop}
\end{figure*}

Without near-infrared data, we must make assumptions
about the properties of the pre-existing stellar populations.
Fig.~\ref{fig:oldpop} shows the range of central-aperture colors
in our sample.  We segregate the sample into
galaxies with substantial ongoing star formation 
({\it left}, EW(H$\alpha) > 10$~\AA) and galaxies with
little ongoing star formation
({\it right}, EW(H$\alpha) < 10$~\AA).  
We expect some contributions from 
triggered star formation even to the right side of the figure
because timescales for blue $B-R$
colors are substantially longer than timescales for 
H$\alpha$ emission.  Despite the contribution from 
triggered bursts, the outer edge of the distribution 
(from interlopers and/or galaxies with no triggered star formation)
provides an upper limit to the appropriate value of \BRo.
For illustrative purposes, we choose a fiducial constant value
$\mBRo = 1.5$ (vertical line in Fig.~\ref{fig:oldpop}b).

Fig.~\ref{fig:OLDoneplot} shows the 78 pair galaxies with measurable
Balmer decrements (hence relatively strong ongoing star 
formation) on the continuous star formation model contours with
$\mBRo=1.5$ for three IMFs and three reddening corrections.  
The points fall in the expected range for the steep
and cutoff Salpeter IMF slopes, based on 
all but the ``traditional'' reddening correction, which 
over-corrects continuum colors. 
For comparison, Fig.~\ref{fig:OLDoneplot_0.9} shows the contours
and data with $\mBRo = 0.9$.  

Although the absolute positions of the points on the contours are
extremely uncertain, for a given \BRo\ the relative positions of the
galaxies in the \BRs/\ewm\ plane are more reliable.  
To the extent that a constant \BRo\ is valid, 
one conclusion from Fig.~\ref{fig:OLDoneplot} is that
the galaxies in our sample with larger \ewm\ have stronger {\it and}
younger bursts of star formation.  The more complex models that follow
confirm this interpretation.

\bigskip
\bigskip
\bigskip

\subsubsection{Constraints on the IMF}

\begin{figure}[tbh]
\epsfxsize=\linewidth
\epsfbox{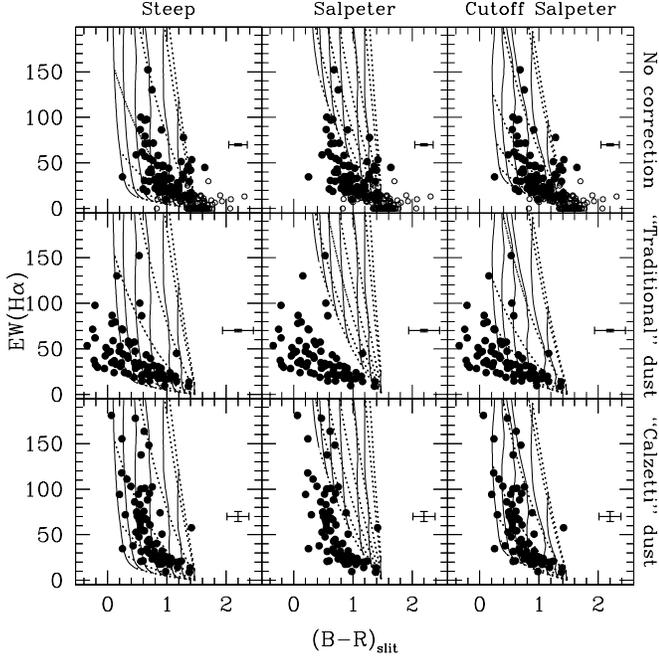}
\caption{Pair data and model contours of constant \strr\ and
t (see Fig.~\ref{fig:contours}).  Our model assumptions are
described in Table~\ref{tab:contours}, except we vary the IMF in the
three columns: ({\it left column}) L99 steep IMF, 
({\it middle column}) L99 Salpeter IMF, and ({\it right column})
L99 Cutoff Salpeter IMF.  We correct the data for extinction
according to three prescriptions for dust (see Sec.~3.2.1):
({\it top row}) no correction (except for Galactic extinction),
({\it middle row}) ``traditional'' correction, and ({\it bottom row})
``Calzetti'' correction.  We omit contour labels; they are the same
as Fig.~\ref{fig:contours}: ({\it solid, left to right}) 
$\msr = 1, 0.8, 0.6, 0.4, 0.2, 0.01$ 
and ({\it dotted, bottom to top}) t $=$ 10$^9$, 10$^{8.5}$,  10$^8$,
10$^{7.5}$, 10$^7$, 10$^{6.5}$, and 10$^6$ years.  The lines on
the right side of the plot show the median errors for the galaxies
with measured Balmer decrements. For dust-corrected
points, errors include propagated errors in the Balmer decrement.
We plot the 112 points with no measured Balmer decrement on the
top line only ({\it open circles}).}
\label{fig:OLDoneplot}
\end{figure}

\begin{figure}[tbh]
\plotone{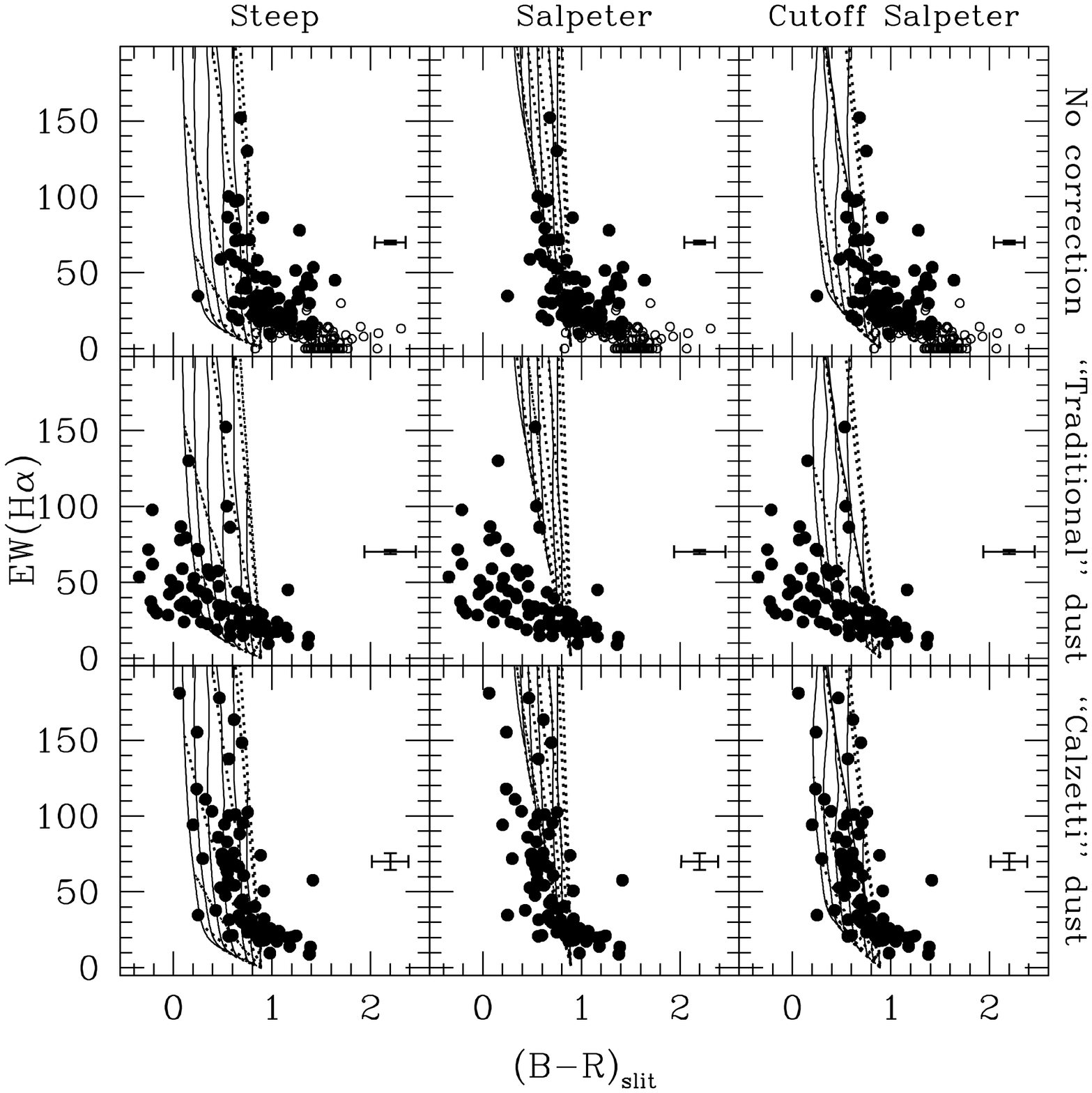}
\caption{Same as Fig.~\ref{fig:OLDoneplot} except
$\mBRo = 0.9$. }
\label{fig:OLDoneplot_0.9}
\end{figure}

For all reddening corrections, most of 
the data fail to overlap the Salpter model contours in
Figs.~\ref{fig:OLDoneplot} and \ref{fig:OLDoneplot_0.9}.  
Most of the data that are bluer 
than the assumed \BRo\ fall below the constant-\BRo\ Salpeter contours, 
corresponding to ages $\gg 10^9$ years.  
Timing arguments alone rule out this possibility for the majority of
the data; thus, if the 
Salpeter slope is correct, one or more of the other assumptions 
are invalid.  Different {\it constant} (galaxy-independent) 
\BRo\ values do not affect the results;
changes in \BRo\ only shift contours horizontally.  
Different star formation histories also fail to solve the problem
convincingly.
Replacing the continuous star formation models 
with short (or instantaneous) bursts yields ages of $\ll 10^7$ 
years for all of the bursts.  Although our data do not strictly 
rule out the instantaneous-burst scenario, interaction timescales 
are $\gtrsim 300$~Myr, or $\sim$50 times the duration of this burst; 
only a tiny fraction of our sample would then be bursting.
This argument does not rule out 
bursty star formation occurring {\it quasi}-continuously.
In a time-averaged sense, a bursty star formation history 
mimics a steeper IMF ---
in the time between bursts, the highest-mass stars are preferentially
removed.  Thus, our data do not strictly distinguish between a
steeper IMF and a shallower IMF with bursty star formation.

For simpler (non-bursty) star formation histories, 
our models and observations suggest either (1) a burst IMF with a high-mass slope 
steeper than the Salpeter IMF or, (2) 
bursts of star formation with a Salpeter-type IMF
that produce smaller changes in the galaxy color.  
In other words, for Salpeter-IMF bursts to work, hence to lie on contours
like those in Figs.~\ref{fig:OLDoneplot} and \ref{fig:OLDoneplot_0.9}, 
the \BRo\ color must shift with each galaxy --- its color must be 
determined not by the 
burst but primarily by the pre-existing galaxy.

Near-infrared colors will help to distinguish between these 
possibilities in the future.
In addition, a different line of argument, available to us now, results
from the anti-correlation reported in BGK between \ew\ and pair separation on the sky,
$\Delta D$.  
In the next section, we adopt different distributions of \BRo\ from
the literature and investigate the relationship between the burst
IMFs and the possible origins of the $\Delta D$~--~\ew\ correlation.

\section{The Origins of the \dd\ -- \BRs\ -- \ew\ Distribution}

The primary result of the spectroscopic study of BGK is 
the dependence of EW(H$\alpha$) and the other emission lines 
on the pair separation on
the sky, $\Delta D$.  The equivalent widths also 
correlate with velocity separation, $\Delta V$.
BGK find that galaxies in pairs with smaller spatial separations have
a much larger range of EW(H$\alpha$), whereas galaxies with
larger separations (up to $\sim$50~h$^{-1}$ kpc) almost 
always have equivalent widths $\lesssim 30$ \AA.  BGK show that
the correlation can be explained
by a scenario in which a burst of star formation begins at a close
pass, continuing and aging as the galaxies move apart.  In this picture,
the dynamical timescales match star formation timescales (for the
steep IMF).  The picture is also broadly consistent with the
hydrodynamical simulations of \citet{MH96,TDTSS02} for progenitor 
galaxies with shallow central potentials (no bulges).  The biggest
uncertainty in the BGK picture is the contribution to EW(H$\alpha$)
from reddening and from old stellar populations in the centers
of the galaxies.  Our analysis of $B$ and $R$ photometry
provides new constraints on this problem.

\begin{figure}[tbh]
\plotone{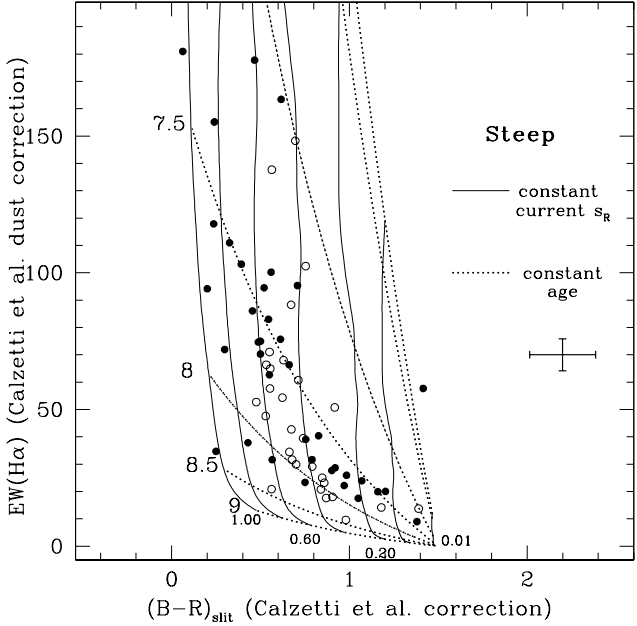}
\caption{Similar to the panel in Fig.~\ref{fig:OLDoneplot} with the
``Calzetti'' dust correction and the steep IMF. The 
points are segregated  based on pair separation, $\Delta D$;  
filled circles are galaxies in pairs
with $\Delta D < 20$~h$^{-1}$~kpc.  Open circles have 
$\Delta D > 30$~h$^{-1}$~kpc.  We omit the galaxies without measured
Balmer decrements.}
\label{fig:CalzMS_dd}
\end{figure}

Fig.~\ref{fig:CalzMS_dd} shows the steep IMF model $\mBRo=1.5$
contours along with our data corrected according to 
the ``Calzetti'' reddening law and segregated based on
$\Delta D$.  We show only galaxies with 
$\Delta D < 20$~h$^{-1}$~kpc (filled circles) or
$\Delta D > 30$~h$^{-1}$~kpc (open circles).   Although 
the results depend on the IMF and the reddening correction, 
the plot suggests that for the single-\BRo\ model,
the bursts in the galaxies with small separations and large \ew\ are typically
both younger {\it and} stronger than the bursts in
the other galaxies, including the galaxies with larger
separations on the sky.  For a constant burst strength
[\strr\ or \stro], the galaxies in pairs with smaller $\Delta D$
tend to have younger bursts.  

In Sec.~3.4, we show that our model fails for
Salpeter-IMF bursts if the progenitor galaxies have a very
narrow \BRo\ range; the Salpeter model contours do not occupy the
appropriate region in \BRs~--~\ew\ space.
These arguments do not, however, rule out Salpeter 
bursts superposed on galaxies
with a wide range of \BRo.  Here, we construct a simple Monte Carlo
simulation to test whether Salpeter- or steep-IMF bursts and correlations
between $\Delta D$ and \strr\ and/or $\Delta D$ and t$_{\rm burst}$ 
can give rise to the appropriate observed ranges of \ew\ and \BRs\
and to the observed $\Delta D$~--~\ew\ correlation 
[and relative lack of $\Delta D$~--~\BRs\ correlation].  
We restrict the comparison to the 72 galaxies with
$\mew > 15$~\AA\ and measured Balmer decrements, also applying this 
\ew\ limit
to the simulated data.

\begin{figure}[tbh]
\plotone{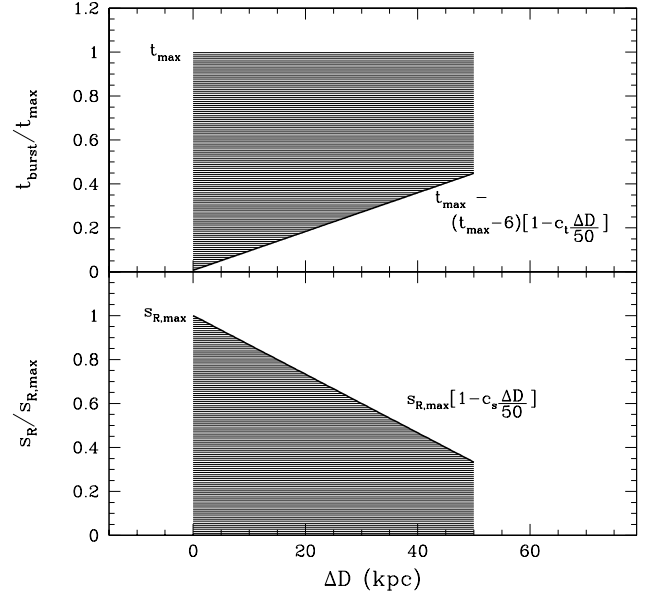}
\caption{An illustration of the simulated ``envelope'' correlations we draw 
from to construct the 72-galaxy simulated data sets that
 we compare with the real data.
For this figure, we use c$_{\rm t} = 4/9$ and c$_{\rm s} = 2/3$ .  The values
we draw for the simulation are uniformly distributed in $\Delta D$; hence they
are not drawn uniformly from the shaded region, but uniformly from the x-axis.}
\label{fig:demo}
\end{figure}

For the simulation, we draw 72 values of \BRo\ from the bulge colors
of \citet{PB96}.  We then draw a value of $\Delta D$ 
from a uniform distribution
as indicated by the data; this uniform distribution represents 
an excess of very tight pairs over a random spatial distribution.  
For each tested maximum burst age and strength, t$_{\rm max}$ and 
s$_{\rm R,max}$,
respectively, we explore the effects of correlations between pair separation,
$\Delta D$, and \strr\ and/or t$_{\rm burst}$.  We test correlations that 
are ``envelopes'', appropriate because $\Delta D$ is only a constraint on the minimum
physical separation of the pair (or compact group).  
For a given $\Delta D$, we draw \strr\ from 0 to s$_{\rm R,max} 
\times \left[1 - {\rm c_s}\frac{\Delta D}{(50\ {\rm kpc})}\right]$.  
This procedure results in a range of \strr\ values from 0 to 
s$_{\rm R,max}$, with the maximum possible
only at $\Delta D = 0$ for nonzero ${\rm c_s}$. Similarly, 
we draw t$_{\rm burst}$ from 
t$_{\rm max} - {\rm (t_{max}-6\ Myr) 
\times \left[1 - c_t \frac{\Delta D}{(50\ kpc)}\right]}$
to t$_{\rm max}$, resulting in a range of
t$_{\rm burst}$ values from 6 Myr to t$_{\rm max}$ Myr, 
with the minimum possible only at $\Delta D = 0$.  The parameters 
${\rm c_s}$ and ${\rm c_t}$ are the slopes of the ``envelopes'' 
and measure the strengths of the $\Delta D$~--~\strr\ and
$\Delta D$~--~t$_{\rm burst}$ correlations, respectively. 
We illustrate these ``envelope'' correlations in Fig.~\ref{fig:demo}.
We compute the ``measured'' \ew\ and \BRs\ directly from the selected \BRo, 
\strr, and t$_{\rm burst}$ using the L99 models with either
the steep or Salpeter IMF.

For several combinations of t$_{\rm max}$, s$_{\rm R,max}$, ${\rm c_s}$
and, ${\rm c_t}$, we repeat this procedure 
1000 times and examine the average.  
We explore the ability of this procedure
to reproduce values close to the averages, \aew\ and \aBRs,
of the data, corrected for extinction via the \citet{CKS94} 
prescription (but not for nebular emission in this case); 
these averages are $\maew = 61$~\AA\ and $\maBRs = 0.71$, 
respectively.  We also explore the relative strengths
of the correlations we produced with this procedure using the 
Spearman rank probability as
our measure of correlation.  The correlation 
strengths for the data are somewhat 
ambiguous.  Without reddening correction or correction for nebular line
emission, the 
$\Delta D$~--~\ew\ and \BRs\ correlations
for these 72 galaxies have strengths of P$_{\rm SR} = 0.0077$ and 
0.033, respectively.
With the ``Calzetti'' correction, which introduces scatter through 
imprecision but also renders \BRs\ and
\ew\ more accurate, the strengths of the Spearman rank correlations 
diminish to
P$_{\rm SR} = 0.059$ and 0.165, respectively.  Because the appropriate values 
of P$_{\rm SR}$ are somewhat unclear, we require only that the average
Spearman rank probabilities satisfy
$0.006 \leq \left<{\rm P}_{\rm SR}\right> \leq 0.06$ for the
simulated $\Delta D$~--~\ew\ distribution, that 
$0.02 \leq \left<{\rm P}_{\rm SR}\right> \leq 0.2$ for the 
simulated $\Delta D$~--~\BRs\ 
distribution, and that the simulated $\Delta D$~--~\ew\ relation 
is, on average,
substantially more significant than the $\Delta D$~--~\BRs\ distribution
averaged over the 1000 simulations.

\begin{figure*}[tbh]
\epsfxsize=3.4in
\epsfbox{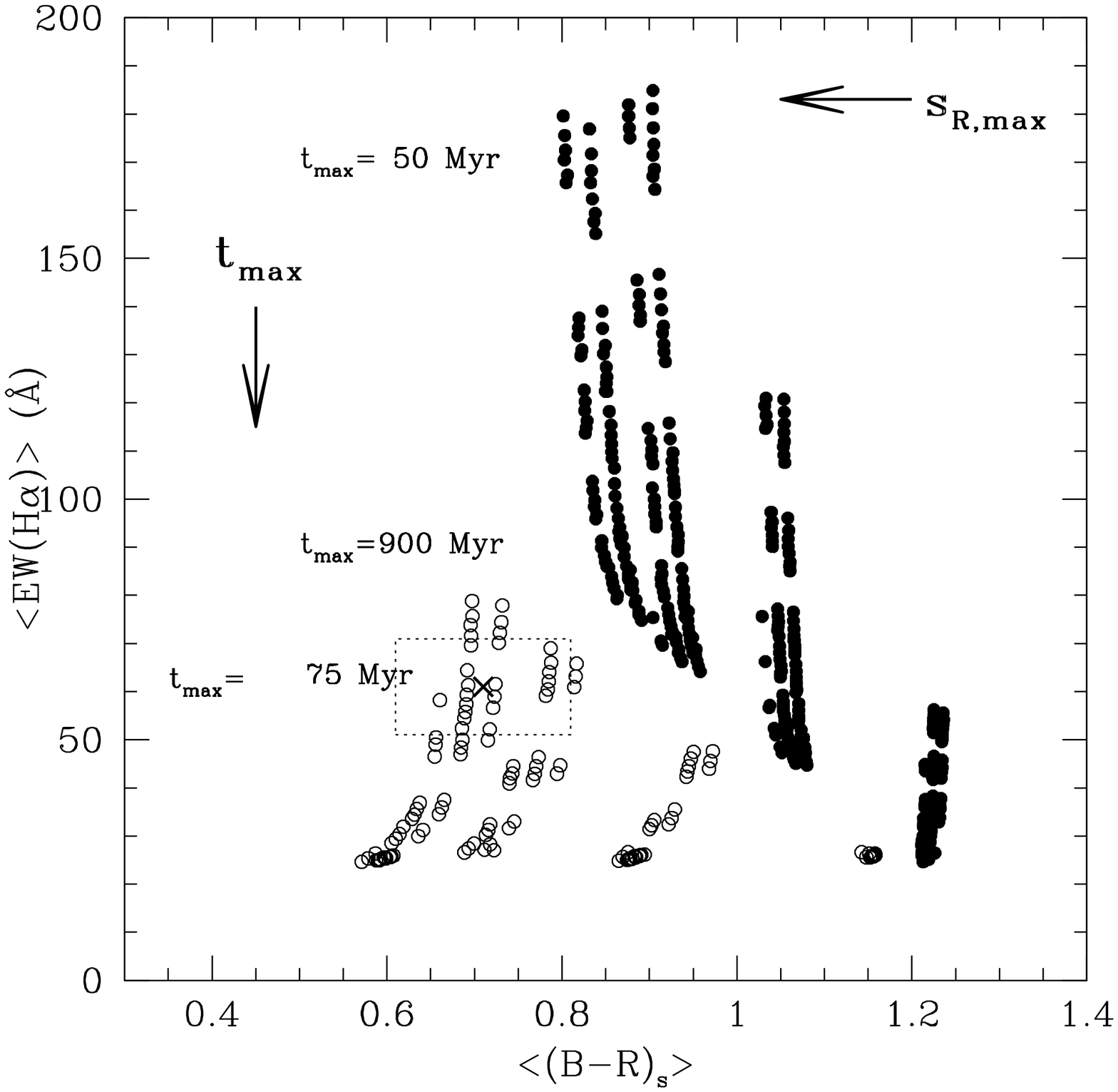}
\epsfxsize=3.4in
\epsfbox{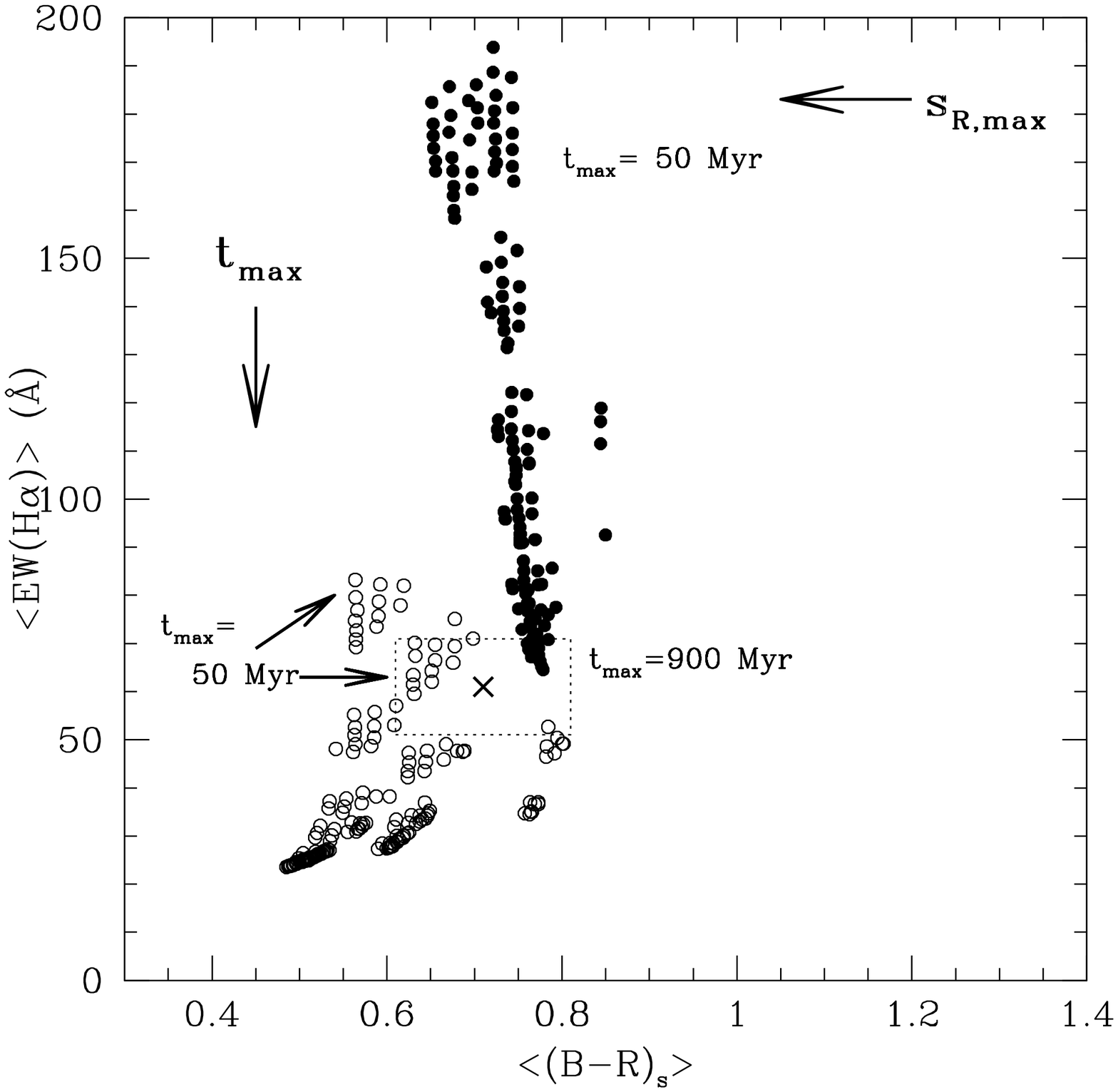}
\caption{Monte Carlo results using the \BRo\ distribution of
\citet{PB96} ({\it left}), and from the NFGS \BRo\ distribution prescription 
({\it right}) for which we draw \BRo\ colors based 
on M$_{\rm B}$ (see text).  
Solid points are Salpeter models; open circles 
show results using the steep IMF.
The box indicates the region surrounding the sample average 
measurements ({\it x}), $\maew = 61$~\AA\ ($\pm 10$~\AA) and 
$\maBRs = 0.71$ ($\pm 0.1$), 
both corrected for reddening via the ``Calzetti'' reddening prescription.
The points show Monte Carlo results for s$_{\rm R,max} = 0.2, 0.5, 
0.8$, and 0.99 ({\it generally left to right}), and 
t$_{\rm max} = 50,100,200,300,500,700$, and 900 Myr ({\it generally 
top to bottom}), omitting
combinations that result in very small \aew.  We add t$_{\rm max} = 75$~Myr 
for the \citet{PB96} colors and steep IMF, to illustrate that the results 
coincide directly with the data.  Considering values of
${\rm c_s}$ and ${\rm c_t}$ from 0 to 1 (in increments of 1/9), we show 
only results that satisfy the Spearman rank correlation properties 
described in the text.  }
\label{fig:monte}
\end{figure*}

Fig.~\ref{fig:monte}a shows \aew\ and \aBRs\ 
resulting from the simulations that 
satisfy the requirements on 
$\left<{\rm P_{SR}}\right>$ for the  $\Delta D$~--~\ew\
and $\Delta D$~--~\BRs\ distributions.
We test only values with t$_{\rm max} \leq 900$~Myr.
For all the tested Salpeter IMF bursts with the \citet{PB96} distribution of
\BRo\ colors, the simulations fail to match the observed
\aew\ and \aBRs.  The most pressing problem is that the 
initial bulge colors are too
red; the \ew\ distribution requires
weak Salpeter bursts for a match while the 
\BRs\ constraint requires strong bursts.
The combination that comes closest to the measured range is that of 
strong (s$_{\rm R,max} = 0.99$) but, on average, older bursts 
(t$_{\rm max} = 900$~Myr).
No simple adjustments to the input 
$\Delta D$~--~\ew\ or \BRs\ correlations will
suffice for t$_{\rm max} \leq 900$~Myr.  
If the \citet{PB96} colors indicate the appropriate distribution of \BRo, 
the Salpeter bursts cannot reproduce the observations.  

Salpeter-IMF bursts on a bluer \BRo\ distribution can meet with more success.
We adopt the distribution of half-light colors 
from the Nearby Field Galaxy Survey of \citet[NFGS]{J00}, restricting to
M$_{\rm B} \leq -17$ (with magnitudes as published but adjusted for 
H$_0 = 70$~km~s$^{-1}$~Mpc$^{-1}$).  Because colors are known
to depend on M$_{\rm B}$, we ensure that the color distribution 
drawn from the NFGS 
is appropriate for a sample with the 
M$_{\rm B}$ distribution of our pairs.  To match 
the distributions, we draw M$_{\rm B}$ from the pair distribution; 
we then assign the galaxy
a \BRo\ that equals the color of an NFGS galaxy selected randomly 
within the $\pm10$ galaxies closest in M$_{\rm B}$.
To take the most conservative approach,
we correct the \BRo\ colors using Balmer decrements drawn from the 
pair sample, to match the $E(B-R)$ in the pair bursts.  This approach 
is conservative as it is appropriate only if the reddening of 
the pre-interaction galaxies equals the 
reddening of the post-burst galaxies.  

Fig.~\ref{fig:monte}b shows the results from
this algorithm.  The Salpeter simulations with strong
s$_{\rm R,max}$ and large t$_{\rm max}$ reach the range of interest
for some simulations with t$_{\rm max} = 700$ or 900 Myr and 
s$_{\rm R,max} = 0.8$.
This maximum timescale is much larger 
that the expected dynamical timescales for
$\sim$50~kpc of separation on the sky.  They may reflect some
contamination in the $\mew > 15$~\AA\ pair population 
from galaxies with older bursts or no 
triggered bursts, but the Salpeter IMF also probably does not result in
blue enough colors for the right range of \ew.
We note that the prescription of 
correcting the \BRo\ colors based on the
measured burst Balmer decrements in our pair sample is probably an overestimate
of the reddening for the pre-existing stellar population.  In addition,
the NFGS does not explicitly eliminate galaxies in pairs and therefore
already contains some triggered bursts.  Finally, the \BRo\
color distribution we seek should include only the central 
few kpc of the galaxies,
whereas the NFGS colors refer to the colors within the effective radius,
frequently larger than the extraction area of our spectra; 
there is more blue disk
contamination in the NFGS colors.  Thus, only a conservatively blue
\BRo\ color distribution reproduces the observations with the Salpeter
IMFs, with what are probably unrealistically long timescales.

For the steep IMF, Fig.~\ref{fig:monte} shows that
several combinations of parameters 
yield average \ew\ and \BRs\ values in the appropriate range,
for either \BRo\ distribution. The best-fitting solutions we find
all have s$_{\rm max} \geq 0.5$ and most have s$_{\rm max} \geq 0.8$.
The solutions work well for a range of ${\rm c_s}$ and ${\rm c_t}$
values; all require relatively strong $\Delta D$~--~\strr\ correlations,
with ${\rm c_s} > 0.33$.  However, the resulting timescales, with
t$_{\rm max} = 50$~--~100 Myr, are unrealistically short.  
We therefore speculate that an IMF steeper than Salpeter, but shallower
than the ``steep'' IMF we employ here, would probably provide the
best match to the data with the most realistic timescales and assumptions.

With the \citet{PB96} bulge colors for the \BRo\ distribution, the 
best-fitting solutions all have $\Delta D$~--~\BRs\ and $\Delta D$~--~\ew\ 
correlations (with ${\rm c_s} \geq 1/3$ and ${\rm c_t} \geq 1/3$ for the
solutions shown).  However, 
for the bluer NFGS \BRo\ colors, 1 of the 13 indicated 
``steep IMF'' solutions 
that fall within the indicated region has ${\rm c_t} = 0$.
Thus, a $\Delta D$~--~t$_{\rm burst}$ correlation is not strictly
required to reproduce the data.

The Monte Carlo simulation is not a rigorous attempt to 
reproduce the data but a simple exploration of the different combinations 
of continuous bursts of star formation which can result in the appropriate 
$\Delta D$~--~\ew\ --~\BRs\ distribution.
We conclude that the result depends on the appropriate distribution of
\BRo\ for the particular sample.  If the \citet{PB96} bulge colors
are appropriate, the models require a high-mass IMF that is substantially
steeper than Salpeter
to explain the observed colors, equivalent widths, and 
$\Delta D$~--~\ew\ correlation using simple star formation histories.  
Adopting much bluer progenitor colors does allow the Salpeter-IMF
bursts to fit our model, but the required colors may be too blue to
be realistic. Our data indicate a burst IMF steeper than Salpeter,
but probably not as steep as $\alpha = -3.3$.

Regardless of IMF, the models explain the observed
$\Delta D$~--~\ew\ correlation only if there is a strong correlation
between pair separation on the sky and the fraction of $R$-band 
flux in the spectroscopic aperture that originates from the new
burst (i.e., a strong $\Delta D$~--~\strr\ correlation).
Most, but not all, of the working models also require a strong
$\Delta D$~--~t$_{\rm burst}$ correlation.
The models which do not require the
$\Delta D$~--~t$_{\rm burst}$ correlation represent a
departure from the models of BGK, which explain the 
$\Delta D$~--~\ew\ correlation with a
$\Delta D$~--~t$_{\rm burst}$ correlation only.  However, as we note, 
a substantial $\Delta D$~--~t$_{\rm burst}$ relationship
is still consistent with the data (and predicted by most of the models
we explore). 
If the correlation still plays some role, then
the interpretation of BGK still holds qualitatively,
but with somewhat shorter timescales than BGK discuss (e.g., by factors of
1.5~--~3).  We require additional data (e.g., near-infrared) to verify the
relative roles of the $\Delta D$~--~t$_{\rm burst}$ and 
$\Delta D$~--~\strr\ correlations.

\section{Conditions that Affect the Strengths and Ages of the Bursts}

Although we lack direct constraints on \BRo\, we briefly
investigate the relationships between third parameters and star formation
more rigorously by inverting \ewm\ and \BRs\ to 
measure the burst age, t$_{\rm burst}$, and strength, \strr.  
Several observational problems complicate
this process.  First, not every galaxy in the
\BRs/\ewm\ plane has a valid solution for every model.  
We omit galaxies with ages $> 10^9$ years.  
For the two galaxies with nominal $\msr > 1$ we
define $\msr = 0.999$ and estimate t$_{\rm burst}$ by increasing
the measured color until a valid solution exists; this procedure
corresponds to moving the points to the right on 
Fig.~\ref{fig:OLDoneplot}, into the edge of the region of valid solutions.
Second, uncertainties in \BRo\ plague this analysis; we focus primarily on
models with constant \BRo\ and models with a linear dependence of
\BRo\ on the log of the rotation speed of the galaxy.
The final observational problem results from
incompleteness in the sample of
galaxies with measured Balmer decrements (hence accurate corrected
colors and equivalent widths); this sub-sample becomes incomplete 
for $\mewm \lesssim 20$~\AA\ (see Fig.~\ref{fig:BalmerDec}).
As a result of this incompleteness, the data are not sampled
uniformly in t$_{\rm burst}$ or \strr.  We explore possible
correlations between third parameters and
t$_{\rm burst}$ and/or \strr\ after truncating the sample 
in accordance with this approximate incompleteness limit.
There are many acceptable combinations of cutoff \strr\ and
t$_{\rm burst}$ that result in different sub-sample sizes.  We
explore two sub-samples: (1) $\msr > 0.3$, 
t$_{\rm burst} < 10^8$ years, and (2) $\msr > 0.4$, 
t$_{\rm burst} < 10^{8.2}$ years.  Table~\ref{tab:subsets} lists the
number of galaxies with valid solutions
in the subsets we consider, including the full sample, the full sample
restricted to galaxies with measured Balmer decrements, and these
complete sub-samples.  The table notes additional prescriptions for
\BRo\ that we describe in Sec.~5.2, and indicates the number of galaxies that
have measured velocity widths, \Vc.

\begin{table*}
\begin{center}
\tablenum{4}
\tablecolumns{6}
\caption{Number of Galaxies with Valid Model Solutions for Subsets of the Data}
\begin{tabular}{llrrrr}
\tableline
\tableline
\multicolumn{2}{c}{Prescriptions} & \multicolumn{4}{c}{Number with Valid Solutions} \\
\tableline
\colhead{} & \colhead{} & \colhead{} & \colhead{With} 
	& \colhead{Complete\tablenotemark{a}} & \colhead{Complete\tablenotemark{a}} \\
\colhead{} & \colhead{} & \colhead{Full} & \colhead{Measured} 
	& \colhead{(t$_{\rm burst} < 10^{8.2}$ yr)}
	& \colhead{(t$_{\rm burst} < 10^{8}$ yr)}  \\
\colhead{ \BRo } & \colhead{Reddening} 
	& \colhead{Sample} & \colhead{H$\alpha$/H$\beta$} 
	& \colhead{($\mstr > 0.4$)}
	& \colhead{($\mstr > 0.3$)} \\
\tableline
\tableline
$\mBRo = 1.5$                                           & ``Calzetti'' & 131 & 77 & 43 & 51\\
$\mBRo = 1.5 +$    & ``Calzetti'' &  68 & 35 & 12 & 16\\ 
\hspace{0.1in} $ 1.5  [\log{V^{\rm c}_{2.2}}-2.5]$\tablenotemark{b}  & & & & & \\
$\mBRo = 1.5 +  $  & ``Calzetti'' &  74 & 39 & 15 & 20\\
\hspace{0.1in} $ 0.75 [\log{V^{\rm c}_{2.2}}-2.5]$\tablenotemark{b}  & & & & & \\
$\mBRo = 1.5$                                           & none         & 114 & 75 & 17 & 28\\
$\mBRo = 1.5 + $   & none         &  57 & 31 &  1 &  5\\ 
\hspace{0.1in} $ 1.5  [\log{V^{\rm c}_{2.2}}-2.5]$\tablenotemark{b}  & & & & & \\
$\mBRo = 1.5 + $   & none         &  62 & 37 &  3 &  8\\ 
\hspace{0.1in} $ 0.75 [\log{V^{\rm c}_{2.2}}-2.5]$\tablenotemark{b}  & & & & & \\
\hline
 \multicolumn{2}{c}{Prescriptions} & \multicolumn{4}{c}{Number with Valid Solutions and Measured \Vc } \\
\hline
$\mBRo = 1.5$                                           & ``Calzetti'' & 72 & 39 & 17 & 21\\
$\mBRo = 1.5$                                           & none         & 61 & 38 & 6 & 11\\
\tableline
\tableline
\end{tabular}
\tablenotetext{a}{Subsample in a ``complete'' region of t$_{\rm burst}$~--~\strr\ space
that includes only galaxies with measured Balmer decrements.}
\tablenotetext{b}{All galaxies in these categories have a measured \Vc.}
\end{center}
\label{tab:subsets}
\end{table*}

\subsection{\strr, t$_{\rm burst}$, and the Orbit Parameters}

\begin{figure}[tbh]
\plotone{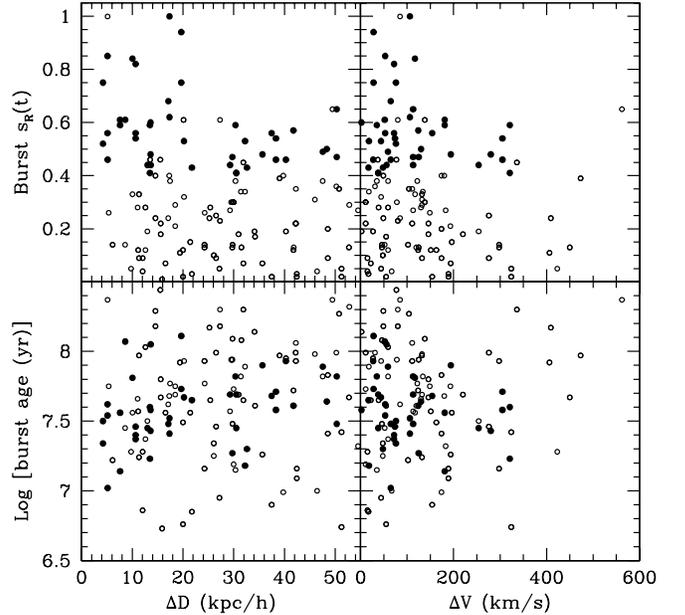}
\caption{Burst parameters (\strr, and age, t,
assuming $\mBRo=1.5$) vs. pair
orbit parameters, $\Delta D$ and $\Delta V$, for the galaxies with
valid solutions only.  Solid points represent the 43 galaxies in
the complete sub-sample (excluding one point with $\Delta D = 76$~h$^{-1}$~kpc
from the left panels); 
they have measured Balmer decrements with 
burst ages $t < 10^{8.2}$ years and strengths $\mstr > 0.4$.
We plot the points without measured Balmer decrements
and/or without valid solutions within
the complete region as open circles.
Within the complete sub-sample, 
there is a tendency for galaxies with smaller values of 
$\Delta D$ to have younger burst ages (P$_{\rm SR} = 0.030$)
{\it and} stronger bursts (P$_{\rm SR} = 0.033$).}
\label{fig:str1}
\end{figure}

Fig.~\ref{fig:str1} shows the burst strength and age
solutions assuming a constant $\mBRo = 1.5$.
We show all the galaxies with 
valid solutions with the steep IMF and the
``Calzetti'' reddening correction.  
We plot burst \strr\ and age, t$_{\rm burst}$, 
as functions of $\Delta D$ and $\Delta V$.  
The filled circles are the
galaxies with measured Balmer decrements 
within the complete sub-sample restricted to
galaxies with $\msr > 0.4$ and t$_{\rm burst} < 10^{8.2}$ years.
The open circles indicate galaxies 
with solutions outside of the complete range and/or galaxies
with no measured Balmer decrement.
The typical formal error in \strr\ is $\sim 0.2$ for 
$0.2 \lesssim \msr \lesssim 0.6$
and is somewhat larger for other strengths.
Errors in burst age are correlated with the burst 
strengths of the galaxies.  For most galaxies, errors
are $\sim$0.2~--~0.3 dex; however, for galaxies with $\msr \lesssim 0.2$
the age errors can span several dex (see Fig.~\ref{fig:OLDoneplot}).

The only visually evident correlation among the 43 solid
points in Fig.~\ref{fig:str1} (complete sample) 
is the absence of very strong bursts with large separations.
However, Spearman rank tests confirm both a 
tendency for galaxies with smaller separations to have younger burst
ages (P$_{\rm SR} = 0.030$) and for galaxies with smaller
separations to have stronger bursts (P$_{\rm SR} = 0.033$).
The other parameter combinations show no correlations,
with P$_{\rm SR} = 0.17$ and 0.76 for the
t~--~$\Delta V$, and \strr~--~$\Delta V$
plots, respectively.  If we instead adopt t$_{\rm burst}=10^8$ years and
$\mstr = 0.3$ as our completeness limits, the significance
of the $\Delta D$ -- t$_{\rm burst}$
correlation increases to P$_{\rm SR} = 0.0019$;
the significance of the $\Delta D$ -- \strr\ 
correlation also increases slightly, to P$_{\rm SR} = 0.012$.
With the inclusion of the galaxies that lack measured
Balmer decrements or that fall in the incomplete parts of the
sample (open circles), a significant correlations
persists between $\Delta D$ and \strr\ (P$_{\rm SR} = 0.038$) while
the significance of the $\Delta D$~--~t$_{\rm burst}$ dependence
drops to P$_{\rm SR} = 0.12$.

\subsection{\BRo, \strr, t$_{\rm burst}$, and the Intrinsic Properties of the Progenitor
Galaxies}

Sec.~4 examines general trends in the star formation
timescales and orbital parameters of galaxies in pairs.
Uncertainties in the properties of the progenitor galaxies,
in the IMF of the triggered bursts,
and in the reddening corrections complicate the
interpretation of these apparent trends.
Even if a theoretically tight relationship exists between the
triggered star formation properties and the positions of the galaxies
in their orbits, we still expect variations in the progenitor 
galaxies (mass, orbit, environment, star formation history, IMF).
These variations add to scatter introduced by
interlopers ($\sim$10\%; see BGK), by galaxies that have not
had a close pass yet, and by
the accessibility of 
only two components of the true spatial $\Delta D$ and one of
the true $\Delta V$.

\begin{figure*}[tbh]
\epsfxsize=\linewidth
\epsfbox{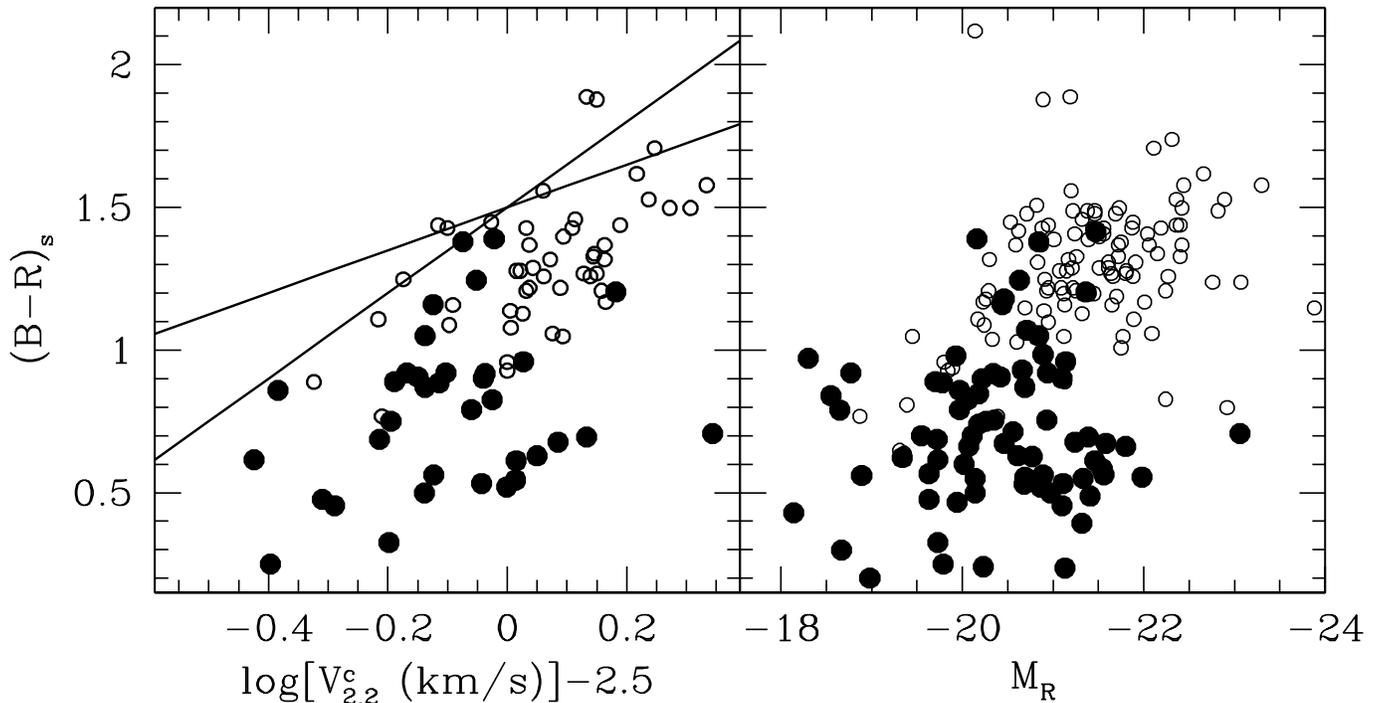}
\caption{$(B-R)_{\rm s}$ corrected for reddening and nebular lines
({\it left}) as a function of rotation curve velocity width 
and ({\it right}) R-band total magnitude for all galaxies 
in the SP sample.  We 
segregate galaxies into groups with measured (solid points) or
unmeasured (open points) Balmer decrements.  The solid lines indicate
the two prescriptions we adopt for \BRo\ as a function of
V$^{\rm c}_{2.2}$.}
\label{fig:color_ada}
\end{figure*}

One powerful tool for sorting out the 
causes of the large scatter we find in Sec.~5.1
is the search for dependence 
of the amount and kind of interaction-triggered star formation
on third parameters \citep[e.g.,][]{TDTSS02}.
In this section, we explore the possible dependence of 
\BRo\ on intrinsic luminosity and rotation speed,
and the relationships between burst
parameters and the velocity widths of the galaxies.

Galaxy colors depend strongly on luminosity, as exhibited by the 
different Tully-Fisher slopes observed for different bandpasses
\citep[see e.g.,][]{T82,TP00,BdJ01}.  However, estimates in the 
literature vary widely and depend on the particular velocity width 
measure.
\citet{TP00} find a difference of only $\sim$0.3 magnitudes per
dex; \citet{BdJ01} quote differences from 0.67 to 1.05 magnitudes
per dex, depending on the reddening prescription.
We concern ourselves only with the colors measured in the
central (spectroscopic) apertures of the target galaxies.
Thus, the extensive Tully-Fisher studies in the literature are
of limited use.  We rely on empirical estimates of
the probable range of variations in color with rotation speed.
Fig.~\ref{fig:color_ada} shows \BRs, corrected for
reddening with the ``Calzetti'' prescription, 
as a function of velocity width,
for galaxies with well-measured rotation curves, 
and as a function of M$_{\rm R}$ for all the
SP galaxies.   [See \citet{B01} for the definition of the
velocity width, which we measure from the major axis rotation curves.] 
The figure clearly shows the dependence of 
color on luminosity or rotation speed; the probable dependence of
\BRo\ on luminosity and rotation speed follows. 

To test the effects of allowing \BRo\ to vary with velocity width,
we adopt the \BRo\ relations drawn in Fig.~\ref{fig:color_ada}a.
The lines correspond to $\mBRo = 1.5 + 1.5 \left[\log{\rm V^c_{2.2}}-2.5\right]$ and $\mBRo = 1.5 + 0.75 \left[\log{\rm V^c_{2.2}}-2.5\right]$.
Using these prescriptions, we find that 
almost all of the correlations in Sec.~5.1 have been reduced in significance.  
The reduced sample size (from 43--51 to 12--20
galaxies; see Table~\ref{tab:subsets}) probably contributes to this 
reduction in significance.
We note that a $\Delta D$~--~t$_{\rm burst}$ correlation persists, 
with P$_{\rm SR} = 0.034$, if we assume 
$\mBRo = 1.5 + 0.75 \left[\log{\rm V^c_{2.2}}-2.5\right]$ and only
accept solutions with t$_{\rm burst} < 10^8$ years and
$\mstr > 0.3$.  
For the ``complete'' subsample with the same limits but with
$\mBRo = 1.5 + 1.5 \left[\log{\rm V^c_{2.2}}-2.5\right]$, 
the correlation disappears, with P$_{\rm SR} = 0.17$.
For the other ``complete'' subsample, these correlations also disappear, 
with P$_{\rm SR} = 0.18$ and 0.33 for the two \BRo\ prescriptions.
Curiously, we also note marginal correlations
between $\Delta V$ and \strr\ for
various restricted sub-samples.

\begin{figure}[tbh]
\plotone{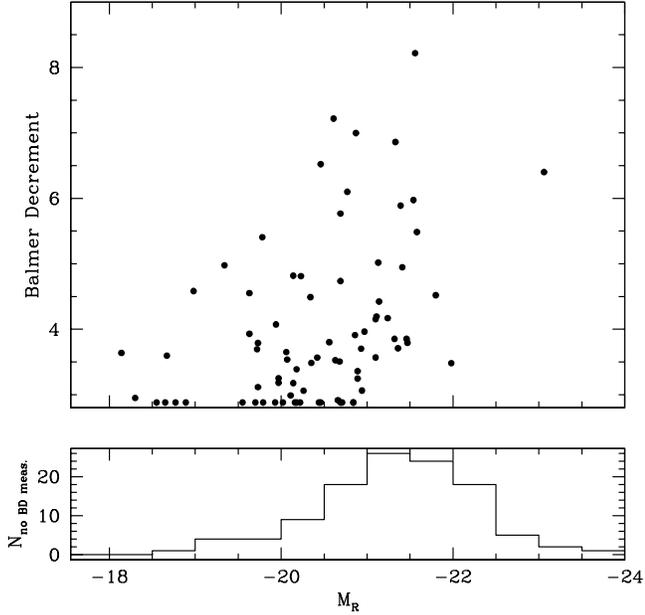}
\caption{Balmer decrement as a function of R-band luminosity for
({\it top}) the 78 galaxies with measured Balmer decrements.  The
data show a tendency for more luminous galaxies to harbor 
more dust (P$_{\rm SR} = 7.1 \times 10^{-5}$).
The histogram shows
the 112 galaxies with H$\beta$ emission too weak to measure
the Balmer decrement.  These galaxies tend to be luminous.
The histogram in Fig.~\ref{fig:BalmerDec} shows their
\ew\ distribution.}
\label{fig:bdmass}
\end{figure}

The metallicities, emission line equivalent widths, and
dust content of galaxies all depend on mass and/or
luminosity \citep[e.g.,][]{T82,SKH89,H98,Car01,Jan01}. 
The required reddening correction thus depends on luminosity
and rotation speed \citep[e.g.,][]{WH96,T98,Su01,Ho01}. 
Fig.~\ref{fig:bdmass} illustrates this very significant dependence in
the SP sample.
The true fractional strengths of bursts of star formation
in these galaxies may also depend on mass.  
Some trends with mass or luminosity appear to
extend to the triggered star formation; in the BGK pair sample, 
the galaxies with the largest measured values of \ew\ 
are the lower-mass, lower-luminosity galaxies.  
\citet{B01} find a shallow slope of the
Tully-Fisher relation for these pairs.

Here, we test for the dependence of the strengths and ages
of triggered bursts of star formation on the masses (rotation speeds)
and luminosities of the progenitor galaxies.  
Timescales for tidally-triggered star formation are
much shorter than a Hubble time.  Thus, even if the frequency of
interaction depends on mass, a large enough,
unbiased set of galaxies
in pairs should show no dependence of burst age on stable
(non-transient) properties like progenitor mass.
The Spearman rank test indicates, as expected, no dependence of
burst age on rotation speed within the 17-galaxy
complete sub-sample (see Table~\ref{tab:subsets})
when we assume a fixed $\mBRo=1.5$
(although a correlation does exist between galaxy R-band
absolute magnitude, M$_{\rm R}$, and t$_{\rm burst}$, 
with P$_{\rm SR} = 0.011$ or 0.031, depending on the 
complete sub-sample).
We introduce very strong false correlations when we
include galaxies in
the incomplete region of t$_{\rm burst}$ -- \strr\ space or
galaxies without measured Balmer decrements; 
with the ``Calzetti'' extinction correction the significance
is P$_{\rm SR} = 0.0014$.  
The complete sample with no dust corrections is too small to test
for a significant correlation. 
The correlation including all the
points in the no-dust-correction scenario is
very strong (P$_{\rm SR} = 3.1 \times 10^{-6}$).
Thus we conclude that restriction of the sample to the complete
sub-sample and the correction for dust are very important 
for eliminating artificial systematic trends (see Fig.~\ref{fig:bdmass}).
Our prescription for changing \BRo\ as a function of \Vc\ 
also eliminates most of the artificial dependence of burst age
on rotation velocity.

Although correlations between galaxy mass and burst age
are almost certainly unphysical, a correlation between
mass and burst strength could be real.
The low-mass galaxies clearly tend to have bluer colors
and higher \ew\ values than the higher-mass
galaxies. 
We investigate the solutions in detail, solving directly for
the burst strengths and ages.  We find some statistical
evidence that the suggested correlation between rotation velocity 
and burst strength exists.  
Assuming a constant $\mBRo=1.5$,
significant correlations exist both within the restricted,
robust, complete sub-sample and for all of the data (including the
galaxies with no measured Balmer decrement).
With a ``Calzetti'' dust correction
the Spearman rank probability is 
P$_{\rm SR} = 0.035$ for the complete sample (restricted
to t$_{\rm burst} < 10^{8.2}$ years, $\mstr < 0.4$; 17 galaxies),
although it is only P$_{\rm SR} = 0.086$ for the other complete
sub-sample. 
The complete no-dust-correction sample is too
small to test for a significant correlation.
However, including all the galaxies with valid solutions,
measured ${\mVc}$, and
no dust correction, the Spearman rank probability 
is P$_{\rm SR} =4.0 \times 10^{-11}$.  
Because these are the prescriptions for which the unphysical burst age-luminosity
and burst age-rotation speed correlations are strong, however,
systematic problems introduced 
by incompleteness and the lack of knowledge of the true \BRo\
almost certainly affect interpretation of the results.

\begin{figure}[tbh]
\epsfxsize=\linewidth
\epsfbox{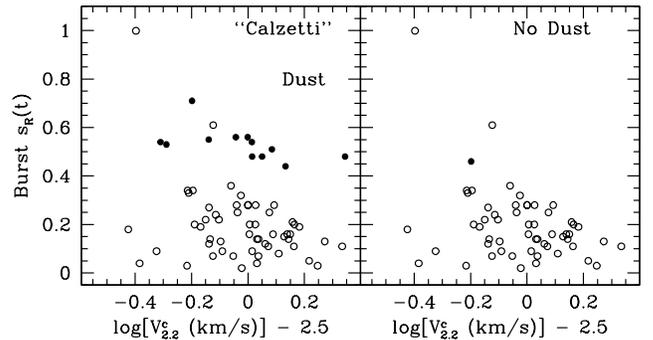}
\caption{Current burst strength for 
$\mBRo = 1.5 + 1.5 \times [\log{V^{\rm c}_{2.2}}-2.5]$
as a function of rotation speed with
({\it left}) a ``Calzetti'' dust correction and ({\it right}) 
no dust correction.
The small number of filled circles have measured Balmer decrements and valid
age and strength solutions within a ``complete region''
($t < 10^{8.2}$ years, $\mstr > 0.4$);
open circles have no measured Balmer decrement 
and/or solutions outside the complete region.}
\label{fig:ada_vs_mstr_vc}
\end{figure}

We test the robustness of the apparent burst strength-rotation speed
correlation by assuming the prescriptions for 
\BRo\ with $\mVc$ that we describe above.  If we assume 
 $\mBRo = 1.5 + 0.75 [\log{\mVc}-2.5]$ and
``Calzetti'' reddening,
the complete sub-sample with 
t$_{\rm burst} < 10^{8.2}$ years and $\mstr < 0.4$
yields a significant
\Vc\ -- \strr\ correlation (P$_{\rm SR} = 0.0023$).
Restriction to the other sub-sample yields no correlation,
however.
For $\mBRo = 1.5 + 1.5 [\log{\mVc}-2.5]$ (see
Fig.~\ref{fig:ada_vs_mstr_vc}),
the \Vc\ -- \strr\ correlation remains 
significant for ``Calzetti'' dust 
(P$_{\rm SR} = 0.011$); again, we observe no correlations within
the second complete sub-sample.  Inclusion of the points outside
the complete sub-sample yields P$_{\rm SR} = 0.49$ (35
galaxies) and including the additional points without
measured Balmer decrements yields P$_{\rm SR} = 0.10$ (for
68 galaxies).
The correlation among all galaxies in Fig.~\ref{fig:ada_vs_mstr_vc}b
is significant, with P$_{\rm SR} = 3.2 \times 10^{-4}$.
We note that of all the combinations we consider in which
\BRo\ depends on \Vc, 
artificial \Vc\ --~t$_{\rm burst}$ correlations are only found using the 
\BRo\ relation with a slope of 0.75 with no dust corrections.

\begin{figure}[tbh]
\plotone{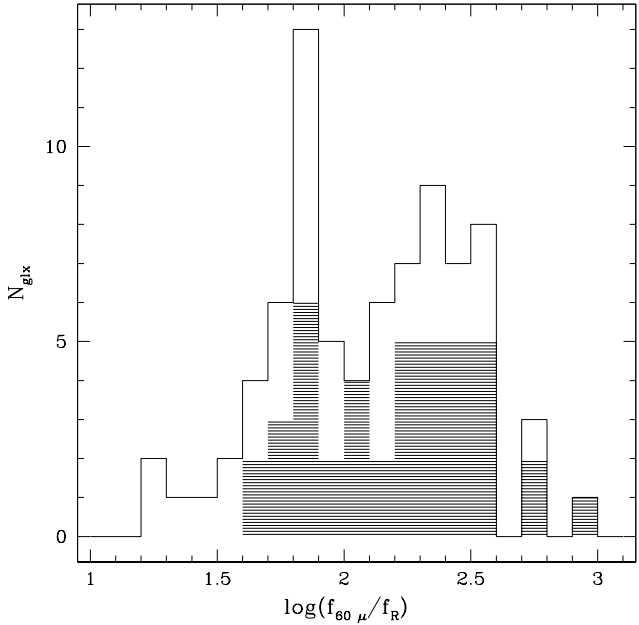}
\caption{For the IRAS psc catalog detections, we plot a histogram
of the log of the ratio of the total IRAS 60 $\mu$m flux within 
3$^{\prime}$ of the galaxy to the $R$-band luminosity.  
We note that in some cases the IRAS flux 
includes flux from both galaxies.  The shaded histogram includes
only the galaxies with measured Balmer decrements (42 galaxies); the main
histogram includes all galaxies with IRAS detections (120 galaxies).
A K-S test indicates only marginally significant 
differences, at best, in the distributions,
with P$_{\rm KS} = 0.12$.}
\label{fig:IRAS}
\end{figure}

The relationship between \Vc\ and the true masses of 
galaxies is not straightforward, especially for interacting galaxies
with centrally-concentrated star formation.
\citet{B01} find several examples of galaxies with truncated rotation curves
that have anomalously small velocity widths.  
When analyzed from the perspective
of another measure of intrinsic galaxy mass, M$_{\rm R}$, the data do not 
present a particularly consistent picture.  For example, with a constant 
$\mBRo = 1.5$, we find no significant burst strength-luminosity correlations 
in the complete subsets of the data.  When we restrict again to galaxies
with measured \Vc\ and assume either of the relations between \BRo\ and
\Vc\, we find a significant \strr~--~M$_{\rm R}$ correlation for one of
the complete sub-sample with t$_{\rm burst} < 10^{8.2}$ and $\mstr > 0.4$.
In addition, we note that there are unphysical M$_{\rm R}$~--~t$_{\rm burst}$ 
correlations, even among the complete sub-samples.  
We do note a strong tendency for galaxies with lower metallicities 
to have stronger bursts.  Although we use solar metallicity for all of
the bursts described here, our tests reveal that the use of lower metallicity
models for the low metallicity galaxies and higher metallicity models for
the higher metallicity galaxies would only enhance this correlation, as
the better-suited models result in stronger bursts at low metallicity and
weaker bursts at high metallicity.  We plan to explore these relationships
in more detail in the future.

\begin{figure}[tbh]
\plotone{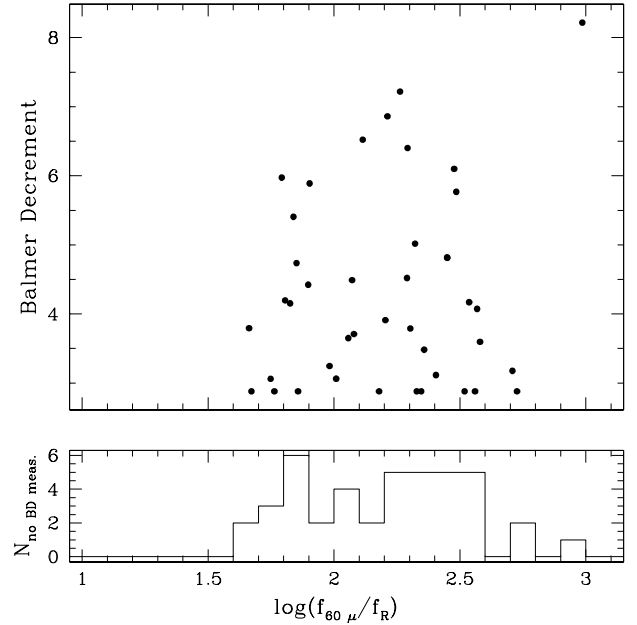}
\caption{The relationship between Balmer decrement and
the 60$\mu$m-to-$R$-band flux for IRAS-detected galaxies in pairs
(see Fig.~\protect \ref{fig:IRAS}.)  The bottom histogram is the
distribution of flux ratios for galaxies without measured Balmer
decrements. }
\label{fig:bdIRAS}
\end{figure}

Our analysis, though far from conclusive, 
suggests that more luminous galaxies may experience 
weaker fractional bursts of star formation.  However, as \citet{H98} point out 
and as Fig.~\ref{fig:bdmass} shows, more massive galaxies are more dust-rich.
Strong dust-enshrouded bursts may hide among the luminous galaxies 
for which we are unable to measure a Balmer decrement.
Hence the apparent conclusion (here and elsewhere) that less massive 
galaxies have higher
luminosity-weighted star formation rates may be an artifact of
inadequate correction for extinction.  

The IRAS point source catalog provides a method of searching for
strong starbursts embedded in the luminous galaxies.
Because the IRAS beam is large, 
we sum all flux within 3$^{\prime}$ of the optical
galaxy, including the flux for both 
galaxies in many close pairs.  Thus, these
ratios are actually upper limits.
Fig.~\ref{fig:IRAS} shows the distribution of the IRAS 
60$\mu$m-flux-to-$R$-band luminosity ratios for the IRAS-detected
sample with measured Balmer decrements (37/78 detected; shaded histogram) 
and the sample as a whole (79/190 detected; open histogram). 
The distributions are at best marginally different, with
P$_{\rm KS} = 0.12$. We find no evidence for extreme hidden
starbursts among the galaxies with 
small \ew\ (hence without measured Balmer 
decrements).

\begin{figure}[tbh]
\plotone{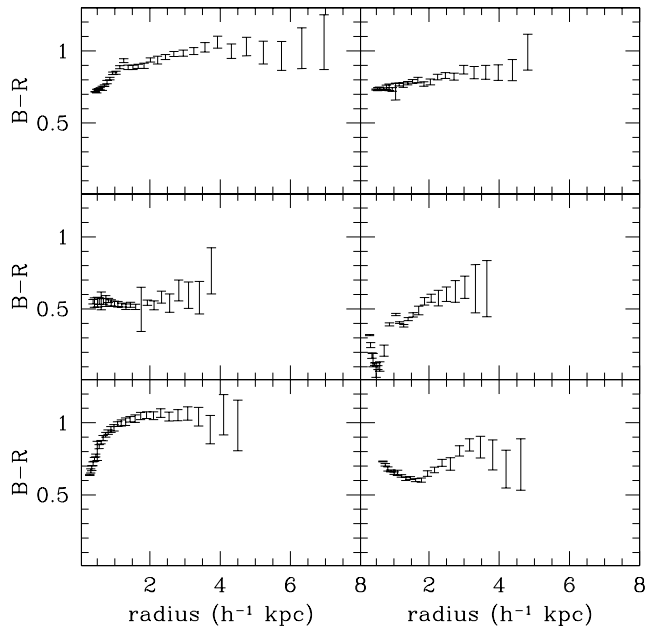}
\caption{Color profiles 
of the 6 galaxies with $\msr > 0.8$ 
(assuming $\mBRo = 1.5$, the steep IMF, and the ``Calzetti'' 
reddening prescription).  These galaxies and many others with strong central 
star formation show blue dips in or near their centers; few of the
galaxies are accompanied by tidal tails, although some companions 
have tails.  Fig.~\ref{fig:strongburst} identifies the galaxies and 
shows $B$-band images.}
\label{fig:BRprofiles}
\end{figure}

\begin{figure*}[tbh]
\begin{center}
\epsfysize=8in
\epsfbox{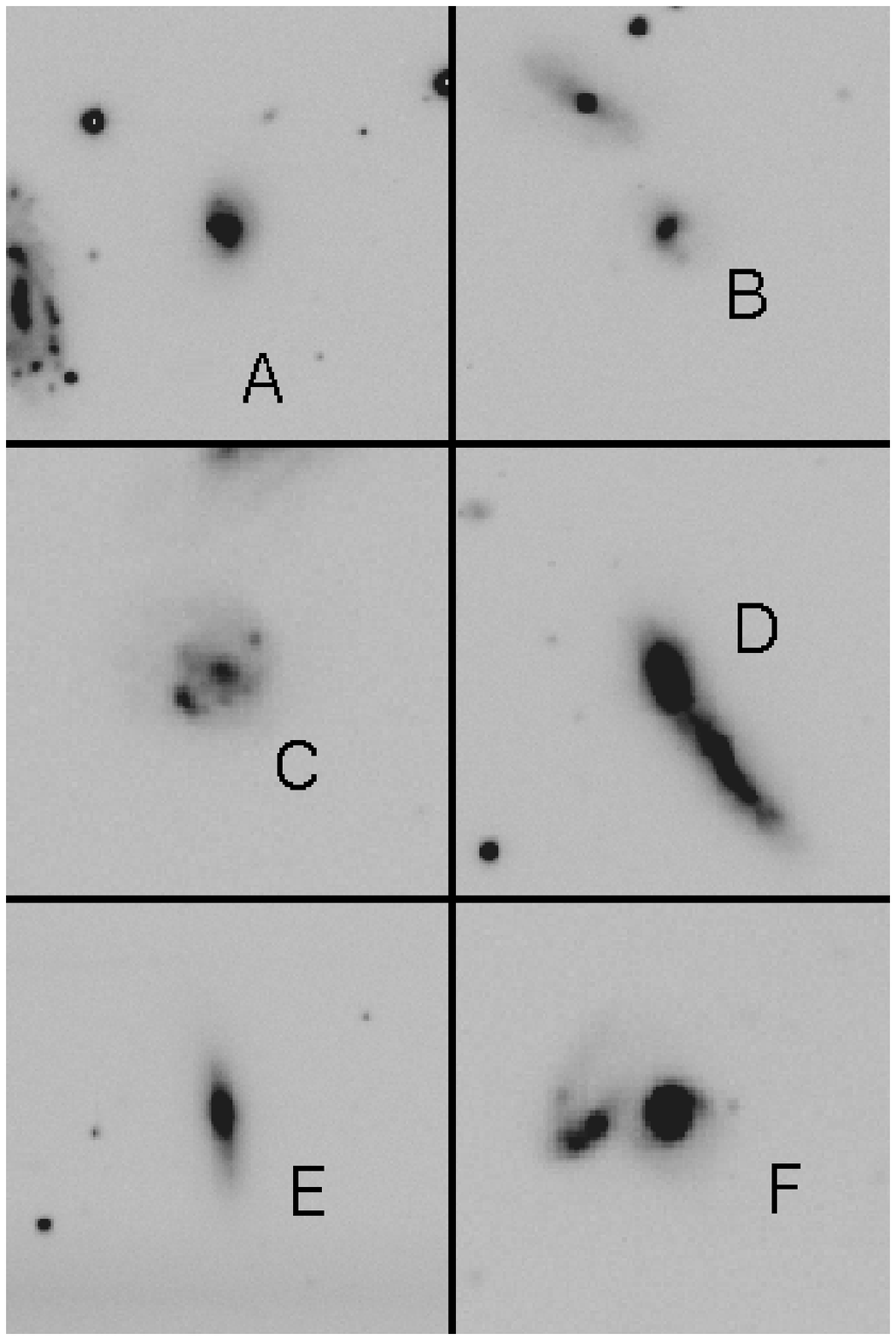}
\caption{$B$-band images of the galaxies in Fig.~\ref{fig:BRprofiles}.
We center the images on the galaxies with nominal $\mstr > 1$.
Galaxies ``A'' -- ``E'' are UGC 312, IC 2338, 
NGC 3991-North, CGCG 132-062, and UGC 12011; they are labeled
in Table~\ref{tab:data}. ``D'' refers to only the Northern galaxy in the close
galaxy pair ({\it middle right}).}
\label{fig:strongburst}
\end{center}
\end{figure*}

We use Fig.~\ref{fig:bdIRAS} to estimate the possible effects
of hidden bursts of star formation on the sample of galaxies
without measured Balmer decrements.  Based on the
rest of the sample, their Balmer decrements are 
probably in the range
of 2.88~--~8, appropriate for the far-IR range in 
our sample.  Thus, the ``Calzetti'' corrections to the 
$B-R$ color and \ew\ may be as large as $\sim$0.83 magnitudes
and a factor of 2.5, respectively, for the galaxies with
the greatest amounts of dust.  Although they
cannot affect the correlations among the galaxies in the
sample with measured Balmer decrements,  more accurate
dust corrections could move the open points in 
Fig.~\ref{fig:ada_vs_mstr_vc} to much
greater strengths, reducing the significance of the
observed \Vc\ -- \strr\ correlation.
Thus, although the IRAS data argue against extremely strong
hidden starbursts,  they do not rule out subtle systematic 
dependences between galaxy luminosity and dust 
content that could give
rise to a false correlation between burst strength and galaxy 
mass.  

In summary, we find some evidence that triggered star formation may
be (fractionally) stronger in lower-mass galaxies.  This result,
if confirmed, would provide one possible mechanism for the stronger
star formation that is typically observed in lower-mass galaxies.
However, we find much less evidence for this trend when we replace
\Vc\ with M$_{\rm R}$ as a measure of the intrinsic mass of a
galaxy.  In addition, known systematic trends in the dust content and 
the stellar populations as a function of rotation speed or luminosity 
of the progenitor galaxies may be affecting
our results.  More accurate 
measures of the dust corrections and pre-existing stellar
population colors are required to explore these issues further.

\section{The Galaxies with Strong Central Star Formation} 

As one product of the analyses we implement in
this paper, we use the \BRs/\ewm\ plane to identify populations
of galaxies with particularly strong bursts or in 
particular stages of interaction.
Fig.~\ref{fig:BRprofiles} shows the $B-R$ color profiles of
galaxies with convincing evidence for strong tidally-triggered bursts.
These six galaxies have $\msr > 0.8$ using the steep
IMF and the ``Calzetti'' dust correction.  We show images of the galaxies
in Fig.~\ref{fig:strongburst}.

The profiles show blue dips in or near the centers;  these dips
are frequent in our sample. By-eye estimates indicate that
22/185 (12\%) of the luminous 
(M$_{\rm B} \leq -17$) non-AGN galaxies have blue central dips in the 
SP sample.  In contrast, the NFGS
contains only 8/147 (5\%) of luminous galaxies with blue central dips.
Because pairs were not
explicitly excluded by the NFGS,
some of the dips in the NFGS sample 
may also result from interactions.

\begin{figure*}[tbh]
\begin{center}
\epsfxsize=6.5in
\epsfbox{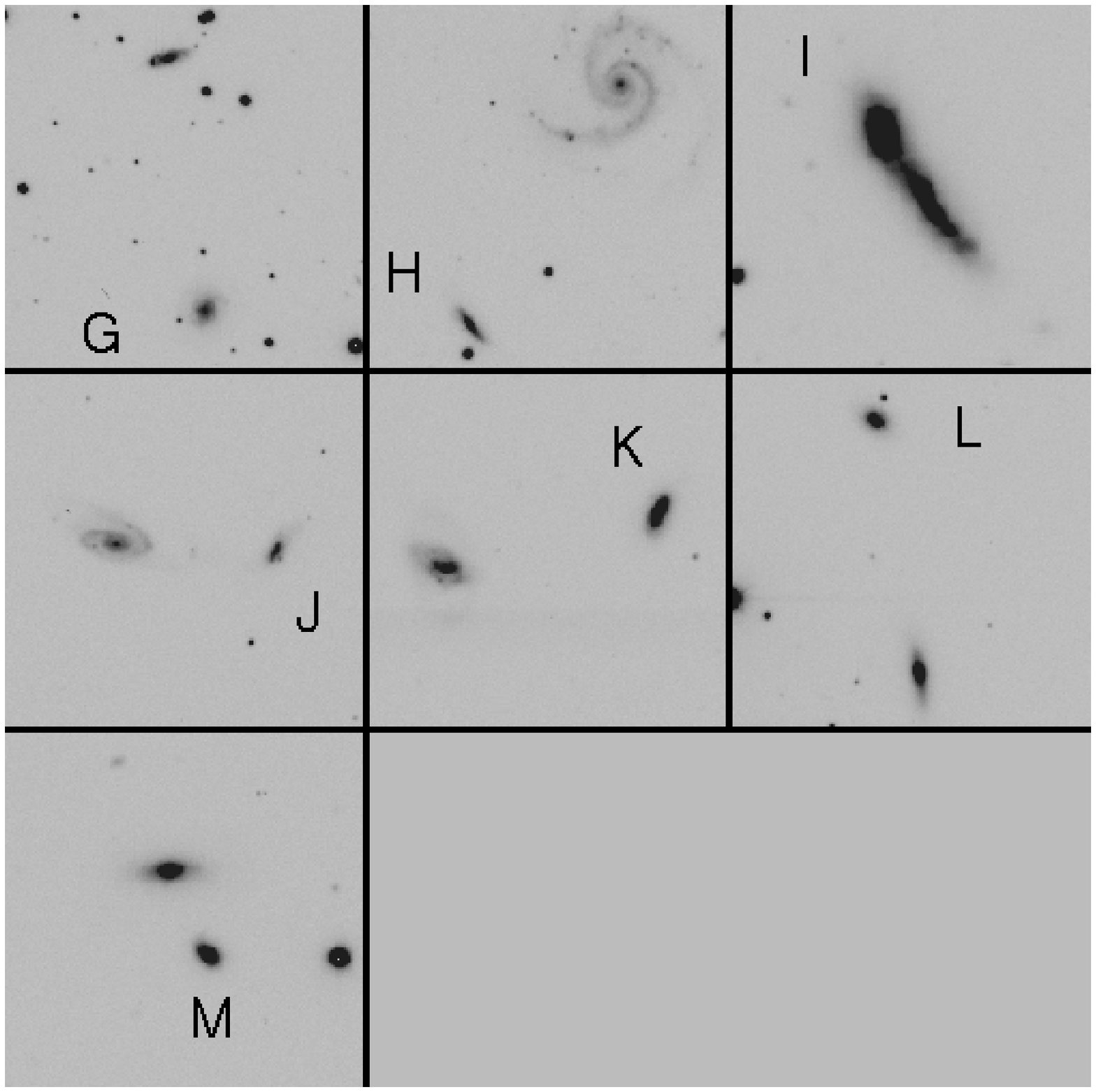}
\caption{Pairs with one member selected based on \strr, \ew,
and $\Delta V$  to be near apogalacticon.  Although some faint features are difficult
to see in this figure, most of the pairs show signs of tidal interaction
and/or lumpy star formation.  
The selected galaxies, ``G'' -- ``M'', are 
CGCG 059-050, CGCG 180-020, NGC 3991-North,
UGC 07085-West, CGCG 217-024, CGCG 132-063, and
UGC 12265-South; they
are labeled in Table~\ref{tab:data}.  (Note that
``D'' in Fig.~\ref{fig:strongburst} is the same galaxy as ``I'' here 
and ``L'' is the companion of ``E''.)}
\label{fig:turnaround}
\end{center}
\end{figure*}

The SPH simulations of \citet{MH96} predict gas infall and 
centrally-concentrated star formation for bulge-less
galaxies in the early stages of 
interactions (see the discussion in BGK).  This centrally
concentrated star formation has implications for the overall
structural parameters of the galaxies, especially at higher redshifts where
low-surface-brightness features are difficult to detect.
\citet{B01} and \citet{BvZ01} study the {\it transient} effects of central 
star formation on emission-line velocity widths and half light
radii.  They find a small number of outliers
to the Tully-Fisher relation in the BGK sample 
with very narrow linewidths for their
luminosities.  Follow-up HI observations with the VLA reveal that
the radio linewidths of these galaxies are much broader; the narrow
optical line widths result from poor
sampling of only the central parts of the rotation curves.
\citet{BvZ01} demonstrate that centrally concentrated star formation
also artificially reduces the half-light radii of galaxies.
Thus measures of linewidths and radii may lead to underestimates of the 
intrinsic masses and sizes of star-forming objects.

Although they exhibit triggered star formation and unusual, lumpy 
morphologies, the target galaxies in Fig.~\ref{fig:strongburst}
have no long tidal tails.  At best, 2/6 have very stubby possible tails.  
Numerical simulations demonstrate that tidal tails are transient and
occur only in prograde interactions \citep{TT72}.  Thus, studies that
identify interactions on the basis of the presence or absence of tidal
tails are incomplete.

Our analyses also provide a method of identifying galaxies that have
already undergone a close pass and are slowing down for a second
pass or merger.  We select galaxies that have been experiencing a
long-lived but ongoing burst of star formation and are near
apogalacticon by choosing galaxies with blue colors
[here, $\mBRs^{\rm c} < 0.75$], moderate \ew\ [20~\AA$ < \mew < $ 40~\AA] 
indicative of a substantial build-up of R-band continuum, and
small velocity separations ($\Delta V < 150$ km~s$^{-1}$).
The morphologies of the chosen pairs suggest that 
this method is nearly 100\% successful at identifying galaxies that have
already experienced a close pass.  We show
$B$-band images of the pairs in Fig.~\ref{fig:turnaround}; one
of the galaxies overlaps with the strong-burst sample
of Fig.~\ref{fig:strongburst}.
Of the 7 galaxies we select with these
criteria, which are members of 7 different pairs, each pair shows one or
more relatively unambiguous indication of a previous close pass (e.g.,
lumpy, unusual morphology or a tidal tail 
either in the blue galaxy itself or in the neighbor).  
A large sample of galaxy pairs near apogalacticon chosen in this manner
would provide kinematic constraints on the dark matter content of galaxies
\citep[e.g.,][]{CS91}.

\section{Conclusion}

We combine $B$ and $R$ photometry with longslit and central spectroscopy 
to explore the strengths and ages of tidally-triggered bursts of
star formation in galaxies in pairs.  We construct a two-population 
description of the pre-existing and triggered stellar populations
near the centers of the galaxies.  Our primary conclusions are:

\begin{enumerate}

\item The most realistic starburst models that describe the data 
statistically include continuous star formation,
a steep (or truncated Salpeter) IMF, and the reddening correction
of \citet{CKS94}.  
Consistency with the Salpeter IMF 
requires a very blue distribution of progenitor-galaxy colors.  

\item In our picture, the models require strong triggered
bursts in some of the pair galaxies; these bursts constitute $\gtrsim$50\%
of the central R-band light.
This star formation leads to a high incidence of galaxies
with blue centers; tidal tails do not always accompany 
these central bursts of star formation.

\item  Our results suggest that the strengths, and most likely
the ages, of triggered bursts of star formation depend
on the galaxy separation on the sky.
Thus, although they do not conclusively verify the model,
our data are consistent with the picture of a 
burst of star formation triggered by a close galaxy-galaxy pass
that continues and ages as the galaxies move apart.  
In this picture, the strongest bursts of star formation occur only in the 
tightest orbits, giving rise to the strength-separation correlations.
The data also suggest a possible correlation between the strength 
of a triggered burst and the rotation speed (mass) of the progenitor galaxy, 
in the sense that low-mass 
galaxies experience stronger bursts 
of triggered star formation.
This result supports the hypothesis that the evolution of galaxies is 
mass-dependent.
Verifying these conclusions requires additional
constraints on the colors and dust content of the pre-existing
stellar populations.

\end{enumerate}

The colors of the pre-existing stellar populations and the
reddening corrections are 
the dominant uncertainties affecting conclusion (3).
More accurate characterization
of the correlations between orbit parameters and star formation properties
in tidally interacting galaxies requires independent 
constraints on the colors of the old stellar populations.
Near-IR observations can provide more direct measurements of 
properties of the older stellar population; 
we plan to report soon on a comprehensive UBVRJHK imaging 
study of a subset of these galaxies in pairs.

\acknowledgments We thank Jan Kleyna, Norman Grogin, and Daniel Koranyi for
useful software and help with the data reduction procedures.
We thank Lisa Kewley for the use of her abundance modeling code, 
and for many helpful comments; we also thank Eric Bell, Liese van Zee, 
Rob Kennicutt, and Rolf Jansen for helpful comments relating to this work.  
We are grateful to Perry Berlind and Mike Calkins for assistance 
with the spectroscopic observations and to Susan Tokarz for assistance
with data reduction.
Support for E. B. G. was provided by NASA through Hubble Fellowship grant 
\#HST-HF-01135.01 awarded by the Space Telescope Science Institute, 
which is operated by the Association of Universities for Research in 
Astronomy, Inc., for NASA, under contract NAS 5-26555.
The Smithsonian Institution supports the research of M. J. G. and
S. J. K.

\clearpage

\end{document}